\documentclass[usenatbib,usegraphicx]{mn2e}
\usepackage{subfigure}
\usepackage{longtable}
\usepackage{graphicx}
\usepackage{graphics}
\usepackage{keyval}
\usepackage{trig}

\def\beq{\begin{equation}}
\def\eeq{\end{equation}}
\def\bey{\begin{eqnarray}}
\def\eey{\end{eqnarray}}

\def\msun{M_\odot}

\def\Msun{M_\odot}

\def\a0{$a_0$}

\title{N-body simulations for testing the stability of triaxial galaxies in MOND}
\author[Wu et al.]{Xufen Wu$^{1}$,HongSheng Zhao$^{1,2}$,Yougang Wang$^{3}$,Claudio Llinares$^{4}$,Alexander Knebe$^{4,5}$\\
$^{1}$SUPA, University of St Andrews, North Haugh, Fife, KY16 9SS, UK\\
$^{2}$Sterrewacht Leiden, P.O. Box 9513, 2300 RA Leiden, Netherlands\\
$^{3}$National Astronomical Observatories, Chinese Academy of Sciences, Beijing, 100012, P.R. China\\
$^{4}$Astrophysikalisches Institut Potsdam, An der Sternwarte 16, D-14482 Potsdam\\
$^{5}$Departamento de F{\'\i}sica Te{\'o}rica, M{\'o}dulo C-XI, Univ.\ Aut{\'o}noma de Madrid, E-28049 Madrid, Spain}
\begin{document}
\maketitle
\begin{abstract}
We perform a stability test of triaxial models in MOdified Newtonian
Dynamics (MOND) using N-body simulations. The triaxial models
considered here have densities that vary with $r^{-1}$ in the center and
$r^{-4}$ at large radii. The total mass of the model varies from $10^8\Msun$
to $10^{10}\Msun$, representing the mass scale of
dwarfs to medium-mass elliptical galaxies, respectively, from deep MOND to quasi-Newtonian gravity. We build triaxial galaxy models using the Schwarzschild technique, and evolve the systems for 200 Keplerian dynamical times (at the typical length scale of 1.0 kpc).
We find that the
systems are virial
overheating, and in quasi-equilibrium with the relaxation taking
approximately 5 Keplerian dynamical times (1.0~kpc). For all systems, the change of the
inertial (kinetic) energy is less than 10\% (20\%) after 
relaxation. However, the central profile of the model is
flattened during the relaxation and the (overall) axis ratios
change by roughly 10\% within 200 Keplerian dynamical times (at 1.0kpc)
in our simulations. We further find that the systems are stable once
they reach the equilibrium state.
\end{abstract}

\section{Introduction}

Elliptical galaxies are often triaxial and appear stable.  A triaxial
equilibrium is non-trivial to build dynamically especially for a system with a
cuspy profile of the light and/or the dark halo. The main objective of this work is to test whether  triaxial models of galaxies are stable in Modified
Newtonian dynamics (MOND, Milgrom 1983). Extensive studies about the
stability of triaxial models have been performed in standard
Newtonian gravity (see below), however, there is no literature on this topic in MOND.

For Newtonian physics, it has been three decades of studies
on constructing a self-consistent model for triaxial
galaxies since Schwarzschild numerically presented the triaxial
Hubble profile in 1979 (Schwarzschild 1979; 1982). Despite its original application to a modified Hubble
profile, the method of Schwarzschild (1979) is still widely used for
testing the self-consistency of various models for the density
distribution in galaxies. For instance, Statler (1987)
showed that the perfect triaxial Kuzmin (1973) profile and the de Zeeuw
\& Lynden-Bell (1985) profile are also self-consistent.
Those models have constant density cores, however,
observations showed that  elliptical galaxies have non-constant
cores (Moller, Stiavelli, \& Zeilinger 1995;  Crane et al. 1993;
Jaffe et al. 1994; Ferrarese et al. 1994, Lauer et al. 1995), i.e. the
surface brightness increases quickly towards the central region of the
galaxies.  Almost all elliptical galaxies have power-law cusps
$\rho \sim r^{-\gamma}$ with $\gamma$ ranging from 1 to 2 for High
Surface Brightness to Low Surface Brightness elliptical galaxies
in the central region. Spherical models with a fixed value of $\gamma$ have been proposed,
e.g. a
$\gamma=2$ model by Jaffe (1983) and a $\gamma=1$ model by Hernquist
(1990). Today such models are rather discussed within a family of
density distributions with $\gamma$ being a free parameter (Dehnen
1993, Carollo 1993 and Tremaine et al. 1994).
In this regard, Merritt \& Fridman (1996)
tested the modified Dehnen profile, \beq\label{den}
\rho (r) ={(3-\gamma)M \over 4\pi abc}{1\over
r^{\gamma}(1+r)^{4-\gamma}}, 0 \leq \gamma <3, \eeq where
$r=\sqrt{({x\over a})^2+({y\over b})^2+({z\over c})^2}, (c\leq b\leq
a)$, $a$, $b$ and $c$ is the long, intermediate, and short axis of
the ellipsoids. They found that triaxial galaxies with central
density cusps $(\gamma=1)$ were in equilibrium and self-consistent in
Newtonian dynamics. The subsequent work by Capuzzo-Dolcetta et al.
(2007) proved that a two-component triaxial Hernquist system,
including a baryonic component plus a Cold Dark Matter (CDM) halo
are also self-consistent.


Modified Newtonian Dynamics (MOND) -- proposed by Milgrom (1983a,b)
as an alternative gravity theory -- was initially designed to
abandon the need for that yet-to-be-discovered dark matter that
(possibly) accounts for as much as 85\% of all matter in the
Universe. MOND, on the other hand, perfectly predicts the rotation
curves of galaxies as well as the Tully-Fisher relation in the
absence of CDM (McGaugh et al. 2000; McGaugh, 2005). Indeed, MOND
successfully matches the observations on a wide range of scales,
from globular clusters (Angus \& McGaugh 2008, in preparation) to
different types of galaxies including dwarfs and giants, spirals and
ellipticals (Milgrom 2007; Gentile et al. 2007; Milgrom \& Sanders
2007; Famaey \& Binney 2005; Sanders \& Noordermeer 2007; Angus
2008a). The development of several frameworks for a relativistic
formulation of MOND (Bekenstein 2004; Sanders 2005; Bruneton \&
Esposito-Far\'ese 2007; Zhao 2007; Skordis 2008) enabled the study
of the Cosmic Microwave Background (CMB) (Skordis et al. 2006; Li et
al. 2008), cosmological structure formation (Halle \& Zhao 2008;
Skordis 2008), strong gravitational lensing of galaxies (Zhao et al
2006; Chen \& Zhao 2006, Shan et al. 2008) and weak lensing of
clusters of galaxies (Angus 2007; Famaey et al. 2007a). As a dynamically selected reference frame, external fields break the Strong
Equivalence Principle (Bekenstein \& Milgrom 1984; Zhao \& Tian
2006; Famaey, Bruneton \& Zhao 2007b, Feix et al. 2008a,b). Consequently, the rotation
curve, escape speed and morphology of galaxies are determined by
both the background and the internal gravity (Famaey et al. 2007b; Wu
et al. 2007, 2008). Despite its great success we need to
accept that even MOND cannot do well without dark matter completely: a
recent study utilizing a combination of strong and weak lensing by
galaxy clusters showed that MOND requires neutrinos of mass $5-7$eV
(Natarajan \& Zhao 2008). And to be consistent with (dark) matter
estimates of galaxy clusters and observataions of the CMB
anisotropic spectrum (as well as the matter power spectrum), MOND
requires neutrino masses of up to $11$eV (Angus 2008b). One theory
capable of accommodating both these requirements is that of a
mass-varying neutrino by Zhao (2008).

In this paper, we utilize a numerical solver for the
MONDian analog to Poisson's equation to study the stability of
triaxial galaxies in MOND. The code named NMODY has  been widely
applied to different problems: it has been applied to study dissipationless collapses, showing that the
end-products are consistent with
several observations (Nipoti, Londrillo \& Ciotti 2006; Nipotti,
Londrillo \& Ciotti 2007a). The code has also been used to study
various important aspects of galaxy formation. Nipoti et al. (2007b) and Ciotti et al. (2007a,b) found
that phase mixing is less effective and the timescale
of galaxy mergers is longer for MOND than for CDM. Recently, Jordi et al. (2009) and Haghi et al. (2009) applied the external fields into the NMODY code and studied the internal dynamics of distant star clusters.
Further, MOND also
produces stronger bars than CDM (Tiret \& Combes 2007), and
hydrodynamical simulations of spherical bulges indicated that there
are tight correlations between bulge mass, central black hole and
stellar velocity dispersion in MOND (Zhao et al. 2008). These
differences and similarities to CDM simulations immediately lead to
the question of the stability of triaxial systems in MOND as
realistic galaxies are not spherically symmetric objects. Wang et al. (2008) recently found that the self-consistency
of a triaxial cuspy centre $\gamma=1$ also exists for MOND. By
extending the original Schwarzschild method and weighting the orbits
during the generation of the Initial Conditions (ICs) for N-body
simulations it is possible to study the stability and future
evolution of these density models
(Zhao 1996). This method proved successful in, for instance,
creating equilibrium ICs for a fast-rotating, triaxial,
double-exponential bar reminiscent of a steady-state Galactic bar
(Zhao 1996) when evolved forward in time using a
Self-Consistent-Field code (Hernquist \& Ostriker 1992).

Whether there are stable galaxy models in MOND is a lacuna in the studies. It is important to 
build stable galaxy models for dynamical studies.
 Here we will expand upon previous work by studying the stability
and evolution utilizing direct N-body simulations. Our target of
study will be an isolated triaxial galaxy with a mild cusp of $\gamma=1$
in the centre within the Bekenstein-Milgrom MOND theory (1984).
We use the same density models applied in Wang et al. (2008),
with total mass ranging from $10^{10}\Msun$ to $10^8\Msun$, respectively, representing medium-mass elliptical galaxies down to dwarf ellipsoidals, which are in quasi-Newtonian to deep MONDian gravity. We generate the ICs utilizing the method outlined in Zhao (1996) and our N-body simulations confirm that these systems are (initially) in quasi-equilibrium and relax on a rather short time scale of only a few Keplerian dynamical times (1.0 kpc) (see below, sub-section~\ref{technicaldetails}). The systems quickly reach a state of equilibrium, consistent with the results of Wang et al. (2008).
The inertial energy changes by less than 10\% and the kinetic energy by less than 20\% during the relaxation process. At the same time, the initial $\gamma=1$ cusps are flattened. After the relaxation, the systems remain stable. We further note that the triaxialities of the systems do not change significantly during 200 Keplerian dynamical times (1.0 kpc). Moreover, the scalar Virial theorem is valid at any time.

\section{Models, Schwarzschild technique, and ICs for N-body}
\subsection{Poisson's equation in MONDian}
The MONDian Poisson's equation can be written as (Bekenstein-Milgrom 1984):
\beq\label{mond}
\nabla \cdot \left[\mu\left({|\nabla \Phi|\over a_0}\right)\nabla \Phi\right] = 4\pi G\rho,
\eeq
where $\Phi$ is the MONDian potential generated by the matter density $\rho$.
For the gravity acceleration constant, we use $a_0=3600$ km$^2$s$^{-2}$kpc$^{-1}$, which is same as adopted by
Milgrom (1983a,b),  Sanders \& McGaugh (2002) and Bekenstein (2006).
The so-called MONDian interpolation function~$\mu$ is approaching 1 for $|\nabla \Phi|>>a_0$ (Newtonian limit) and $\mu \to {|\nabla \Phi|\over a_0}$ for $|\nabla \Phi|<<a_0$ (deep MOND regime),
and the gravity acceleration is then given by $\sqrt{a_0 g_N}$, taking the place of the Newtonian acceleration $g_N=\nabla \Phi_N$ at the same limit. For our simulations we chose the 'simple' $\mu$-function in the form of (Famaey \& Binney 2005; Zhao \& Famaey 2006; Sanders \& Noordermeer 2007)
\beq
\mu(x)={x\over 1+x} .
\eeq
Furthermore, we use the density distribution given by Eq.~\ref{den}, choosing $\gamma=1$, and $M$ being the total mass of the system.
For our simulations, we choose the ratios $a:b:c=1:0.86:0.7$, with $a=1$kpc.

\subsection{Initial Potential}\label{static}
A very important step in our calculation is to solve Poisson's equation in MOND (cf. Eq.~\ref{mond}).
This is achieved via numerical integration utilizing the N-body code NMODY (Ciotti et al. 2006; Nipoti et al. 2007a) on a spherical grid of coordinates $(r, \theta ,\psi)$.
To this extent we applied a grid of $256\times64\times128$ cells. Note that we yet do not evolve the system forward in time; we simply extract the potential of our (static) density distribution and use it for the Schwarzschild method detailed below.

\subsection{Schwarzschild technique}\label{schw}
Since Schwarzschild (1979, 1982) pioneered the orbit-superposition method to construct self-consistent
models of galaxies, this technique has been widely applied in dynamical studies
(e.g., Zhao 1996; Rix et al. 1997; van der Marel et al. 1998; Kuijken 2004; Binney 2005; Capuzzo-Dolcetta et al. 2007).
The essence of this method is to sample phase-space with a large number of orbits. Properly assigning weights
to different orbits can then give rise to the mass distribution we are interested in.

Specifically, let $N_{\rm orbits}$ be the total number of orbits and $N_{\rm cells}$ be the number
of spatial cells (both will be specified below). For each orbit $j\in N_{\rm orbits}$, we count the fraction of time, denoted by
$O_{ij}$,  that it spends in each of the cells $i\in N_{\rm cells}$. The occupation
number $W_j$ for each orbit $j$ is then determined by the following set of
linear equations

\beq\label{weights}
\sum_{j=1}^{N_{\rm orbits}} W_jO_{ij}=M_i,   ~~~~~~~~~~ i =1,...,N_{\rm cells} \ ,
\eeq

\noindent
where $M_i$ is the mass in cell $i$ expected from the given mass
distribution. We have checked the sum $\sum_{i=1}^{N_{\rm cells}} O_{ij}$ equals unity for each orbit. There are now various choices of how
to actually solve Eq.~\ref{weights}: liner programming
(Schwarzschild 1979; 1982; 1993), Lucy's method (Lucy 1974; Statler
1987), maximum entropy methods (Richstone \& Tremaine 1988; Statler
1991; Gebhardt et al. 2003), or least-square solvers (Lawson \&
Hanson 1974; Merritt \& Fridman 1996; Capuzzo-Dolcetta et al. 2007).
We chose the least-square method (cf. Wang et al. 2008).

\begin{figure}{}
\begin{center}
\resizebox{6.5cm}{!}{\includegraphics{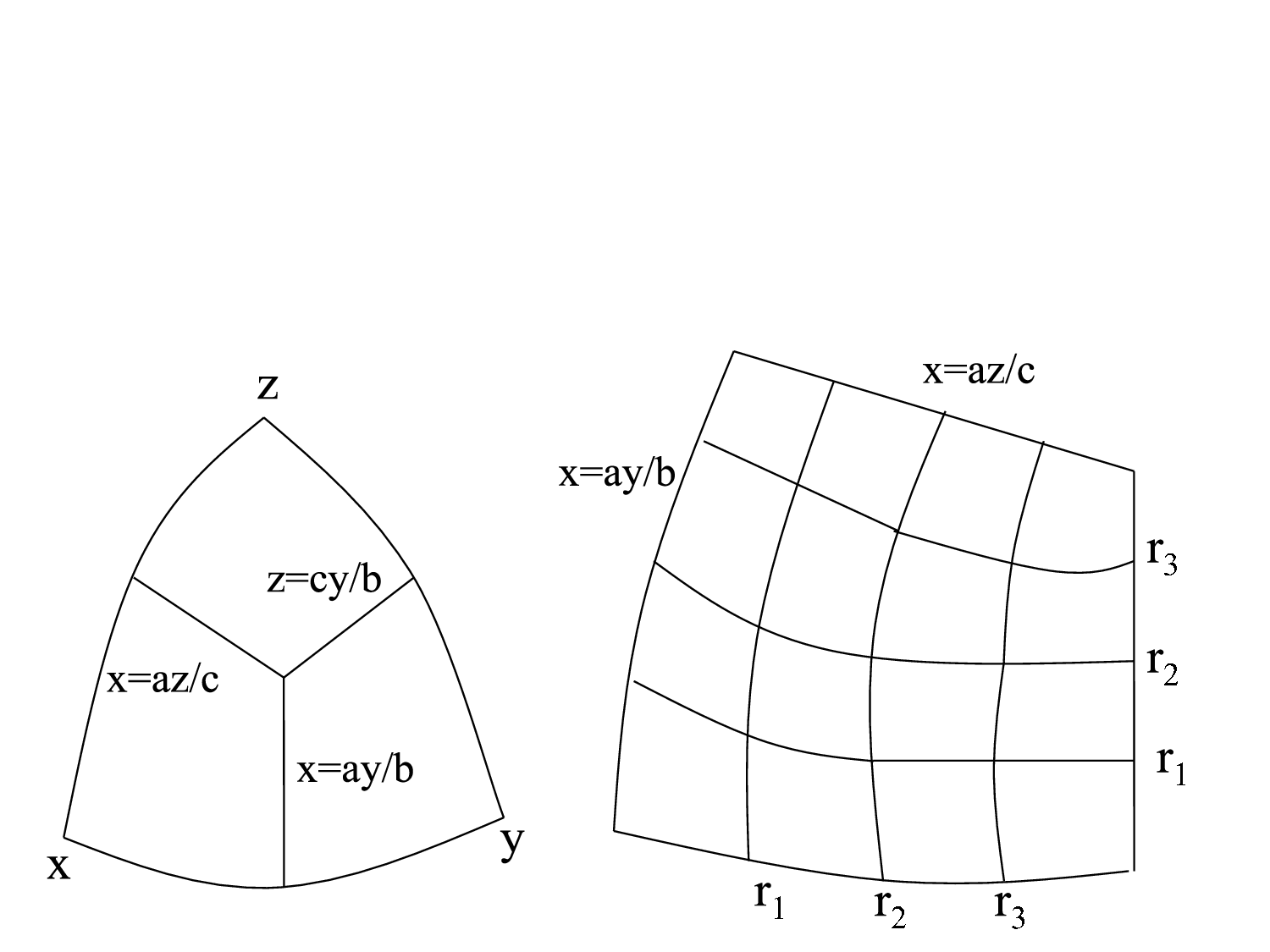}}\vskip 0.50cm
\resizebox{6.5cm}{!}{\includegraphics{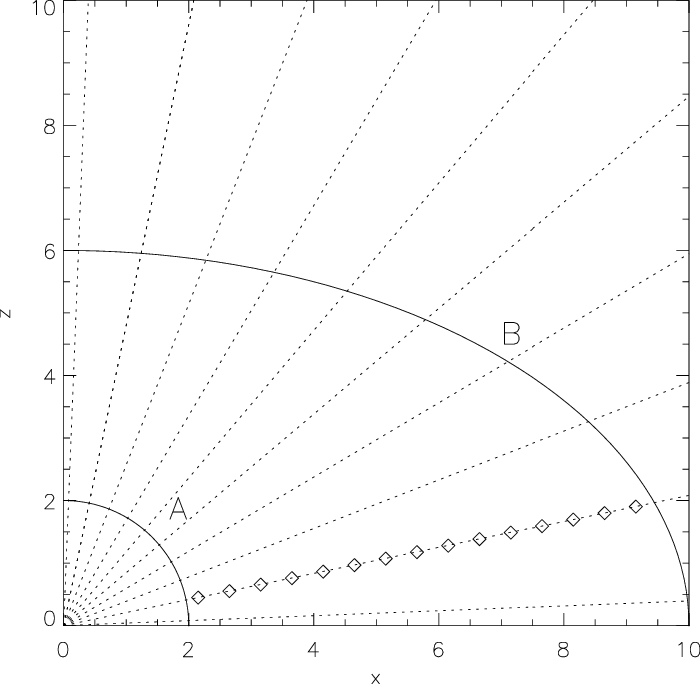}}\vskip 0.0cm
\makeatletter\def\@captype{figure}\makeatother
\caption{The first octant is divided by planes $z=cx/a$, $y=bx/a$ and $z=cy/b$ (upper left panel) into three parts. Each part is subdivided by planes $ay/bx=1/5$,~$2/5$,~$2/3$ and $az/cx=1/5$,~$2/5$,~$2/3$ into 16 cells (upper right panel). In the lower panel, curve A is the circle of the minimal radius of 1:1 resonant orbit at the energy $E_k$, curve B is the zero-velocity surface of the energy $E_k$. There are 10 dotted lines $x=z\tan \theta$ divide the values of $\theta$ from $2.25^\circ$ to $87.75^\circ$. The 15 diamonds equally divide the radius into 16 parts. The diamonds are the initial positions from which the $x-z$ orbits are launched.}\label{cells}
\end{center}
\end{figure}

Because of the symmetry of the mass distribution specified by Eq. (1),
it is sufficient to only consider mass cells in the first octant in our
analyses. Following Merri \& Fridmann (1996), we divide the first octant into
cells of equal masses, i.e. the first octant of each initial system is divided into 21 sectors by 20 shells, where every sector has
the same amount of mass inside. The first octant is further divided into
three parts by the planes $z=cx/a$, $y=bx/a$ and $z=cy/b$ (see Fig.~\ref{cells}, upper left panel).  Each of these three parts
is sub-divided into 16 cells by the planes
$ay/bx=1/5$,~$2/5$,~$2/3$ and $az/cx=1/5$,~$2/5$,~$2/3$ (see Fig.~\ref{cells}, upper right panel). 
Therefore, the total number of cells in the first octant is $16\times 3\times 21=1008$. However, we only consider the innermost 912 cells (the inner 19 sectors) when solving Eq.~\ref{weights} since only a few orbits supply densities in the outermost sector
of grid cells; we simply discard the cells of the outermost sector.

In a spherical system, the total energy and the three components of
the angular momentum are integrals-of-motion. However, in a triaxial
system only the energy remains constant (Merritt 1980;
Valluri \& Merritt 1998; Papaphilippou \& Laskar 1998). We
therefore use the 7/8 order Runge-Kutta algorithm (Fehlberg 1968)
for orbital calculations, with 100 orbital times (see below
Section~\ref{technicaldetails}) for the full integration time of
each orbit. We followed Schwarzschild (1993) and Merritt \& Fridman
(1996) in assigning initial conditions from one of two sets of
starting points (cf. also Wang et al. 2008): stationary orbits with
zero initial velocity, and orbits in the $x-z$ plane with $v_x=v_z=0$,
and $v_y=\sqrt{2(E-\Phi)}\neq 0$ in the first octant.  Note that
there is quite a large number of non-symmetric orbits that will
lead to artifacts in the procedure if only being considered in the
first octant. To circumvent this problem and to keep the symmetry of
the system, we reflect the orbits from the boundaries of each octant.
Note that by our method the computational workload is not increased as would be the case when calculating eight octants.

As mentioned above, there are $16\times 3$ cells in each sector. To obtain enough orbits for the library, we sub-divide the cells by the midplanes of the cells once again, i.e., the midlines of each grid as seen in the upper right panel of Fig.~\ref{cells}; these midlines equally divide the cell at $x=az/c$ and $x=ay/b$. Thus we have $4\times 16\times 3 = 192$ sub-cells in each sector. The central points on the outer shell surfaces of the sub-cells are the launching points. Hence there are 192 stationary starting orbits in each sector.
The total energy on the $k$th sector is defined as the outer boundary shell $E_k$, and for the stationary orbits inside the $k$th sector this amounts to $E_k=\Phi(x,y,z)$. 
For the orbits launched from the $x-z$ plane the initial energy is $E_k=\Phi(x,0,z)$, as shown in Fig.~\ref{cells} (lower panel). This figure further shows that the radius of the inner shell (marked as curve A) is the minimal radius of 1:1 resonant orbits (x:y), and the outer shell (marked as curve B) is the zero velocity surface. We define 10 lines satisfying $x=z\tan \theta$ where $\theta$ lies within the range $2.25^\circ$ to $87.75^\circ$. Along the radial direction, we equally divide the radius between two boundaries into 16 parts with 15 points, where those 15 points are the initial positions for the orbits launched from the $x-z$ plane. Hence there are 150 $x-z$ plane starting orbits.

\begin{figure}{}
\begin{center}
\resizebox{8cm}{!}{\includegraphics{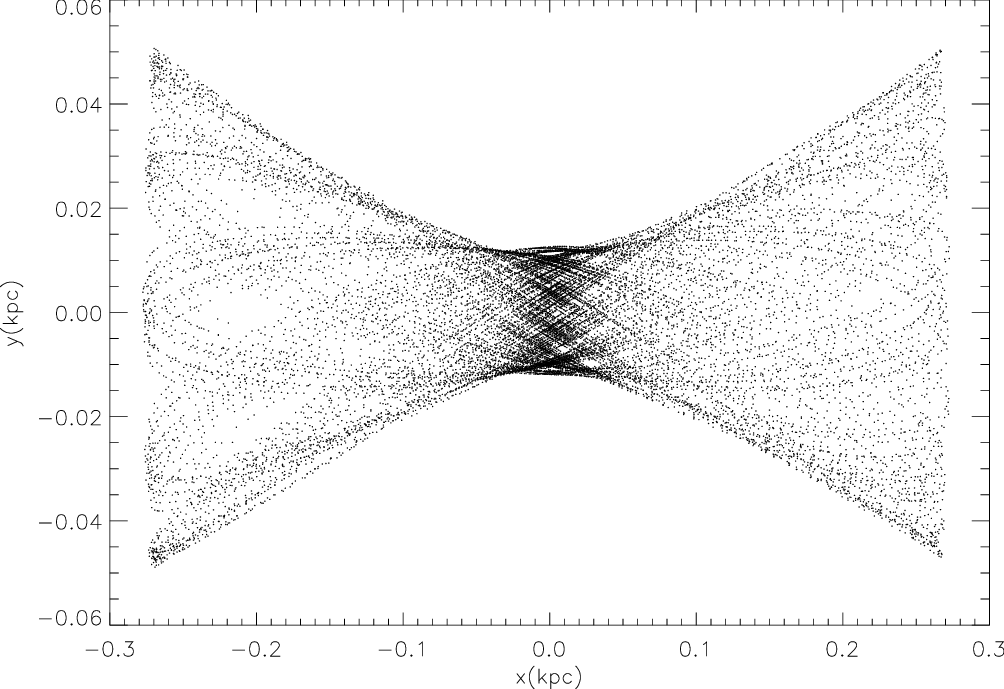}\includegraphics{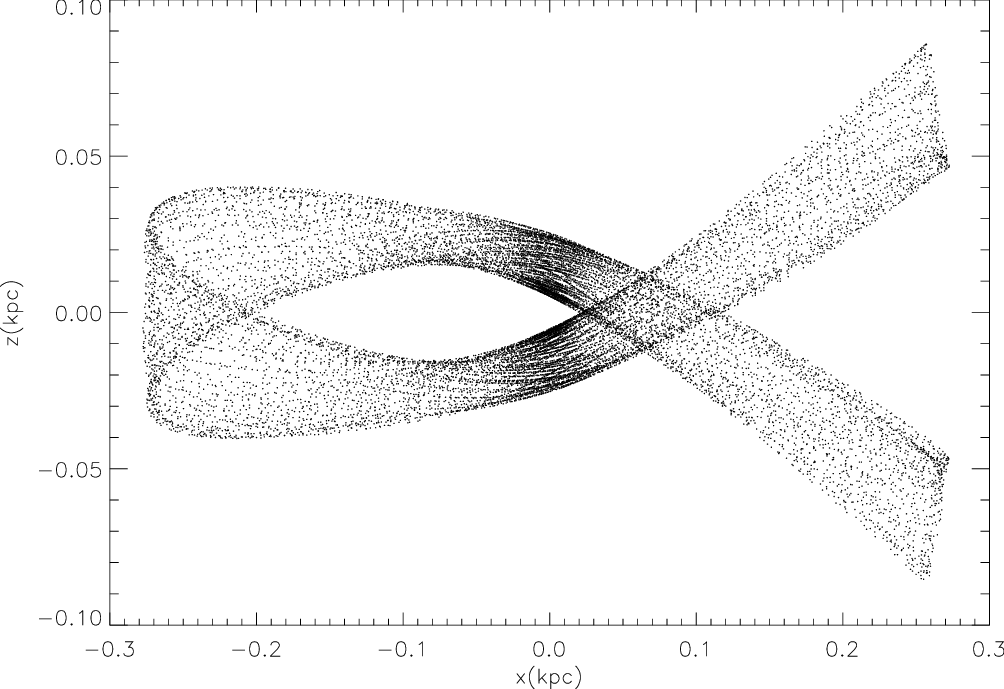}}\vskip 0.00cm
\resizebox{8cm}{!}{\includegraphics{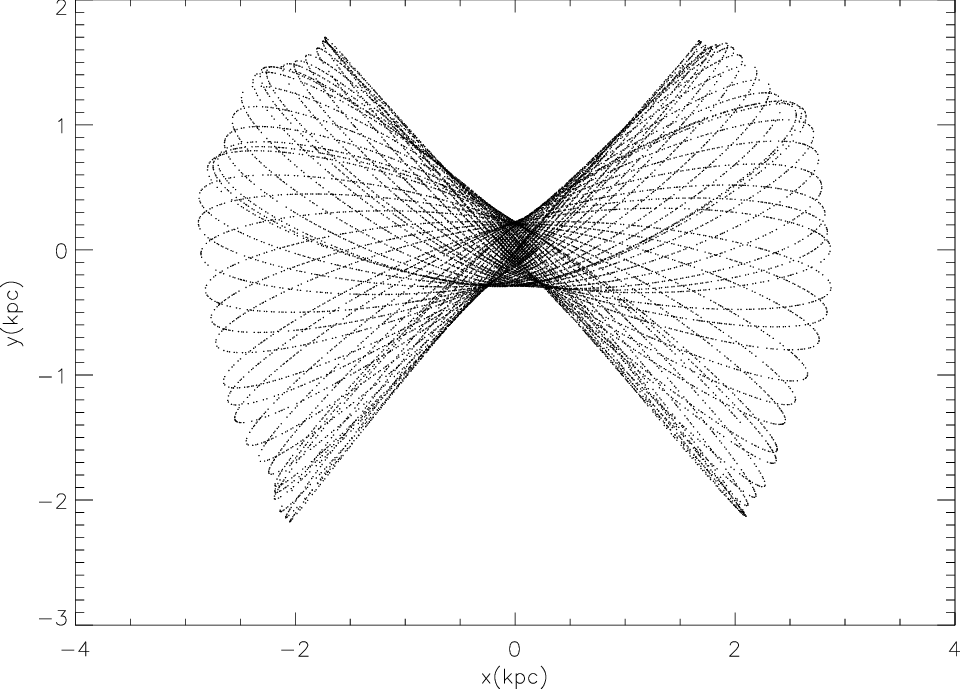}\includegraphics{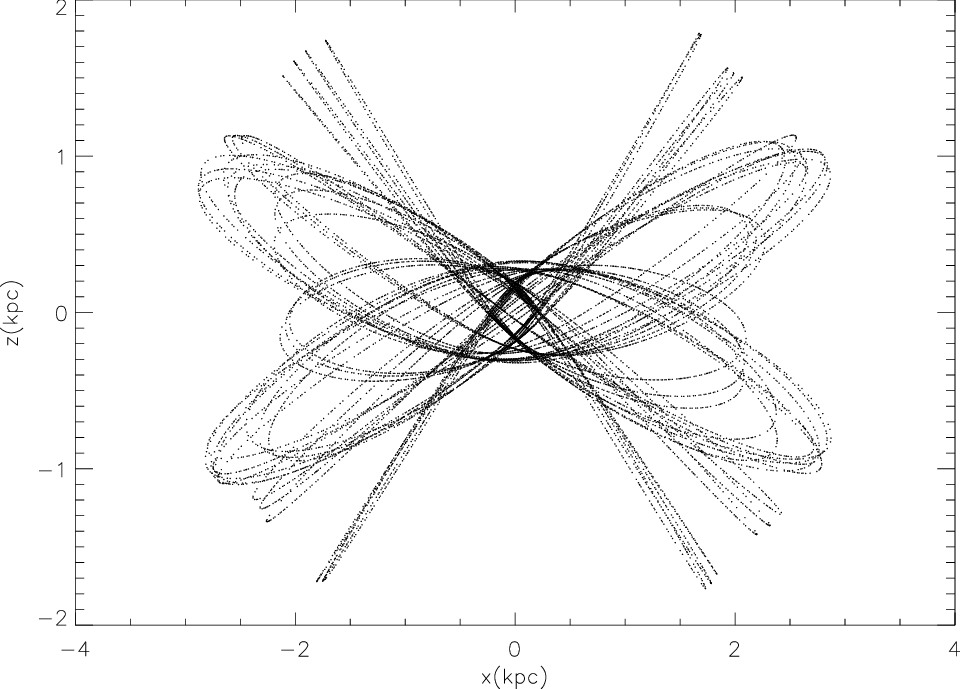}}\vskip 0.00cm
\resizebox{8cm}{!}{\includegraphics{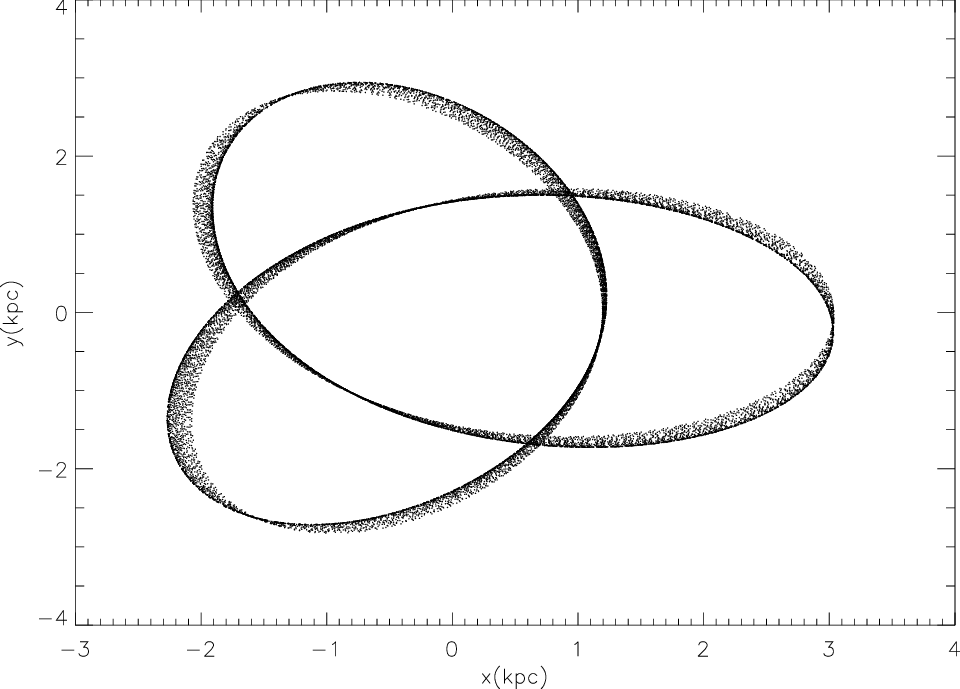}\includegraphics{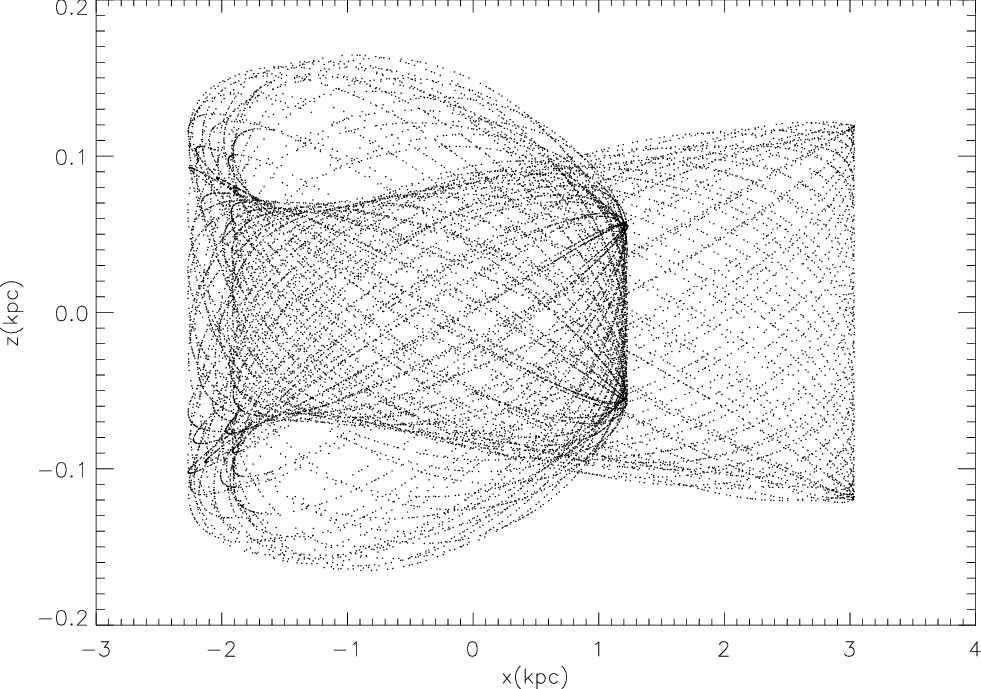}}\vskip 0.00cm
\resizebox{8cm}{!}{\includegraphics{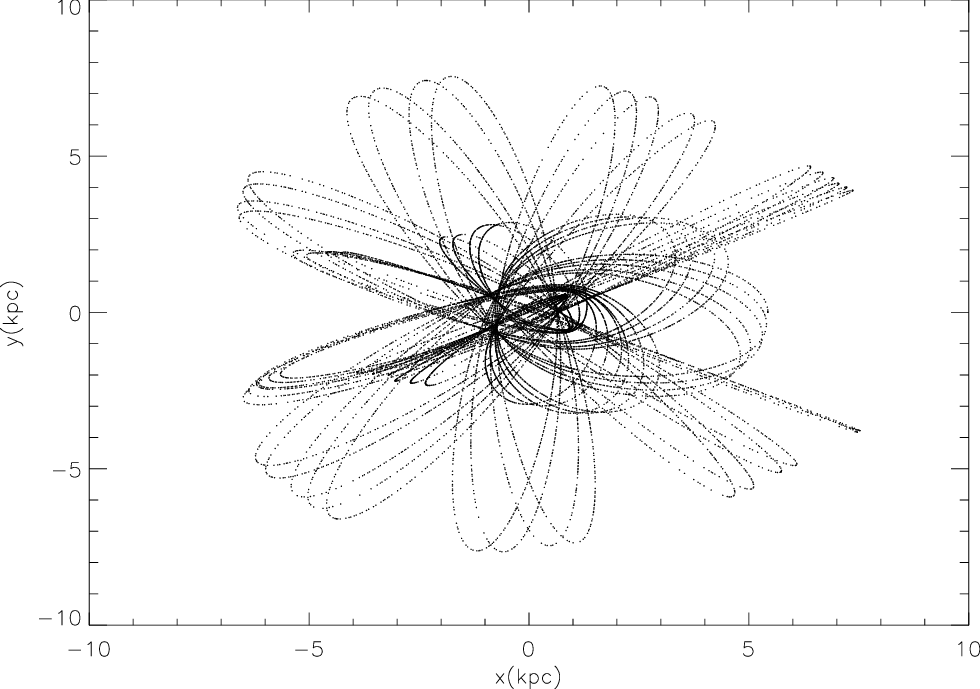}\includegraphics{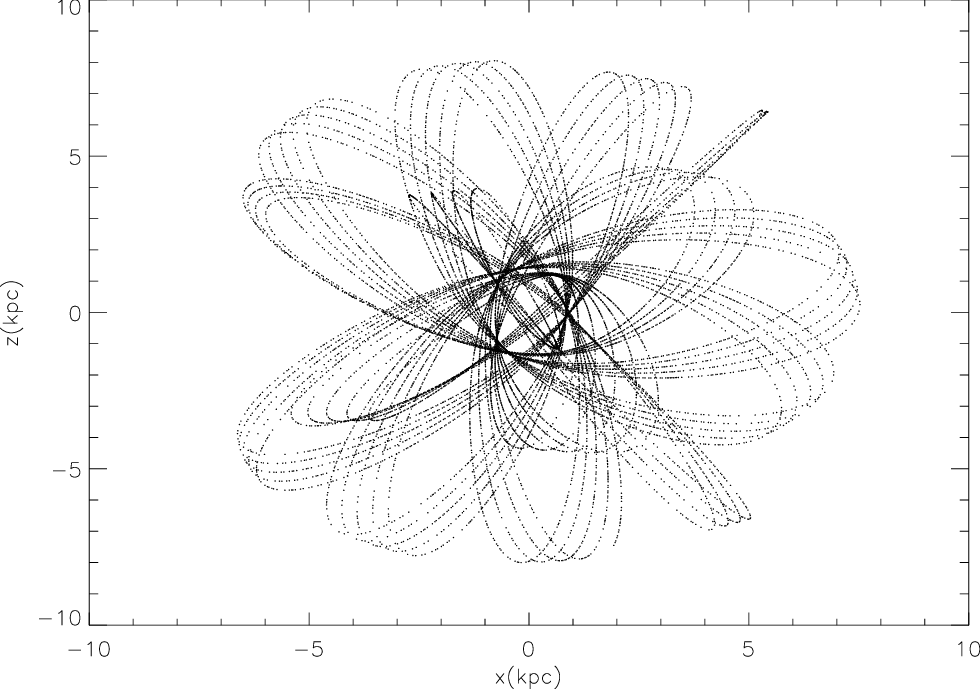}}\vskip 0.00cm
\vskip 0.00cm
\makeatletter\def\@captype{figure}\makeatother
\caption{The asymmetric non-zero weights orbits in the orbit library. The \textbf{upper four panels}: stationary starting orbits and the \textbf{lower four panels}: x-z plane launched orbits. All the \textbf{left panels} are orbits projected on x-y plane and \textbf{right panels} are projected on x-z plane.}
\label{orbits}
\end{center}
\end{figure}

In summary, we have 192 stationary and 150 $x-z$ plane orbits in each sector amounting to a total of $N_{\rm orbits}=6840$ and we use $N_{\rm cells}=16\times 3\times 19=912$ cells for the generation of our orbit library. The energy in each sector is a constant which equals the potential energy on the outer shell surface. As a result, the energies for the systems can be considered 'quantized' with each system having 19 'energy levels'.
In Fig.~\ref{orbits} there are some examples of asymmetric orbits: the upper four panels are stationary starting orbits, and the lower four panels are orbits launched from the $x-z$ plane. In order to generate equilibrium Initial Conditions for the $N$-body simulations to be presented in Section~\ref{simulations}, we need to symmetrize the orbits by using 'mirror particles'.

\begin{figure}{}
\begin{center}
\resizebox{6.5cm}{!}{\includegraphics{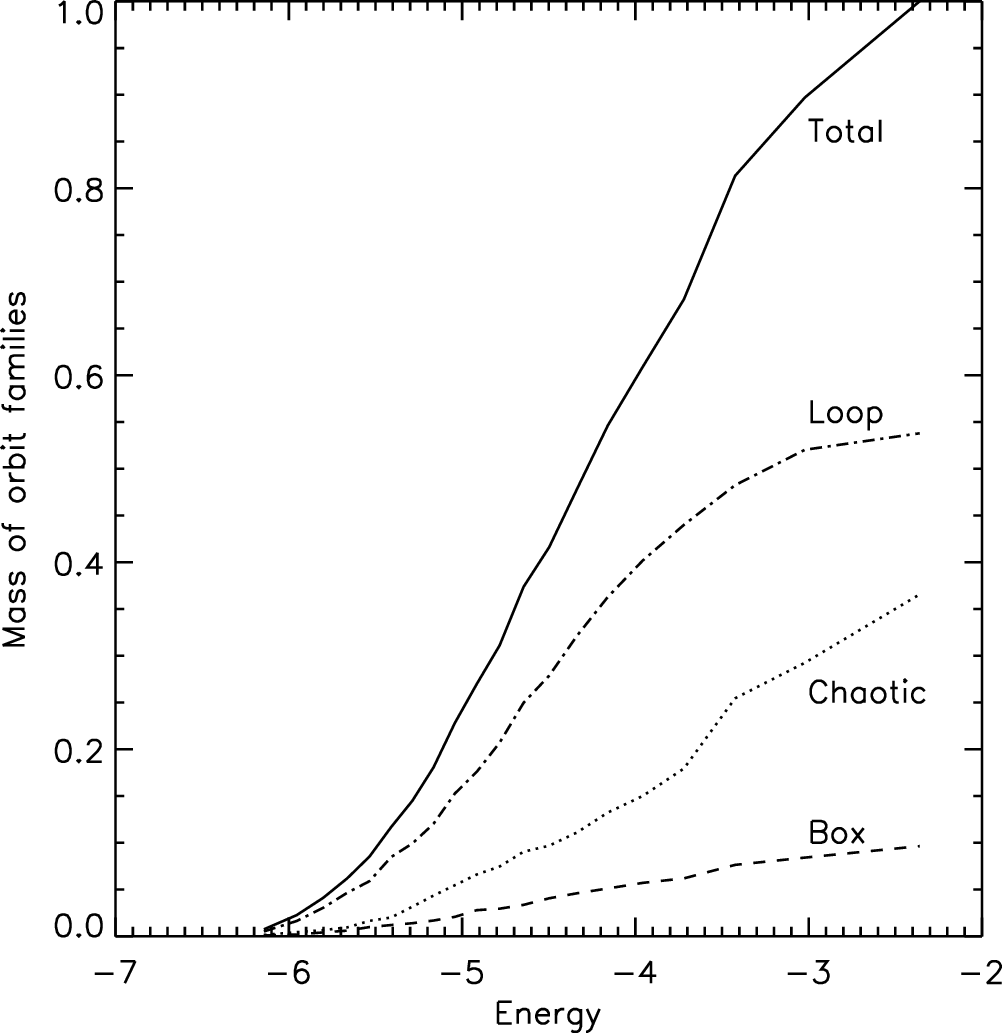}}\vskip 0.5cm
\makeatletter\def\@captype{figure}\makeatother
\caption{The accumulated of energy distribution of different orbit families in the intermediate model with a total mass of $M=10^9\Msun$. The horizontal axis is dimensionless energy of ${E\over GM/1.0kpc}$, and vertical axis is the integration of mass as a function of energy. The dashed, dotted, dot-dashed lines are for box, chaotic and loop orbits. The solid line is for all of the orbits.}\label{orbitene}
\end{center}
\end{figure}

Finally, in Fig.~\ref{orbitene} we show the integration of mass as a function of energy for the model with a mass of $10^9\Msun$. There we find that more than half of the mass stems from loop orbits with chaotic orbits contributing more than $1/3$ of the mass; box orbits therefore do not play an important role in the model.
We further like to note that the self-consistency of the model in MOND has been examined in Wang et al. (2008), and Antonov's third law was applied to check the stability of the models initially. However, it is unknown whether or not Antonov's third law is also valid in MOND so far. The most direct way to check for the stability and investigate the evolution N-body of the system is by means of N-body simulation to be elaborated upon in the following sub-section.

\subsection{ICs for N-body systems}
In order to study the stability and evolution of the systems, we
need to convert the orbits into an N-body model. According to
Zhao (1996), the number of particles $n_j$ on the $j$th
orbit is proportional to the weight of the orbit, i.e., for an $\mathit{N}$
particle system there are $W_jN$ particles on the $j$th orbit. Here we sample the particles on the $j$th orbit isochronously  at $t_j= {T_j\over n_j}\times (i+0.5), i=0,1,2,...,n_j-1$ where $T_j$ is the total integration time of the $j$th orbit. To this extend we interpolate the positions and velocities  from the  6-dimensional output data of the Schwarzschild orbits. 
We generate
\beq n_j=W_jN\eeq
 particles on the j$\mathit{th}$ orbit and symmetrize the particles in  phase-space. The remaining particles are kept as our Initial Conditions of the N-body system.

\section{N-body simulations in MOND} \label{simulations}
All results presented in this section were obtained by evolving our systems forward in time with using the N-body particle-mesh code NMODY (Ciotti et al. 2006; Nipoti et al. 2007a).

\subsection{Technical Details} \label{technicaldetails}
\begin{figure}{}
\begin{center}
\resizebox{6.5cm}{!}{\includegraphics{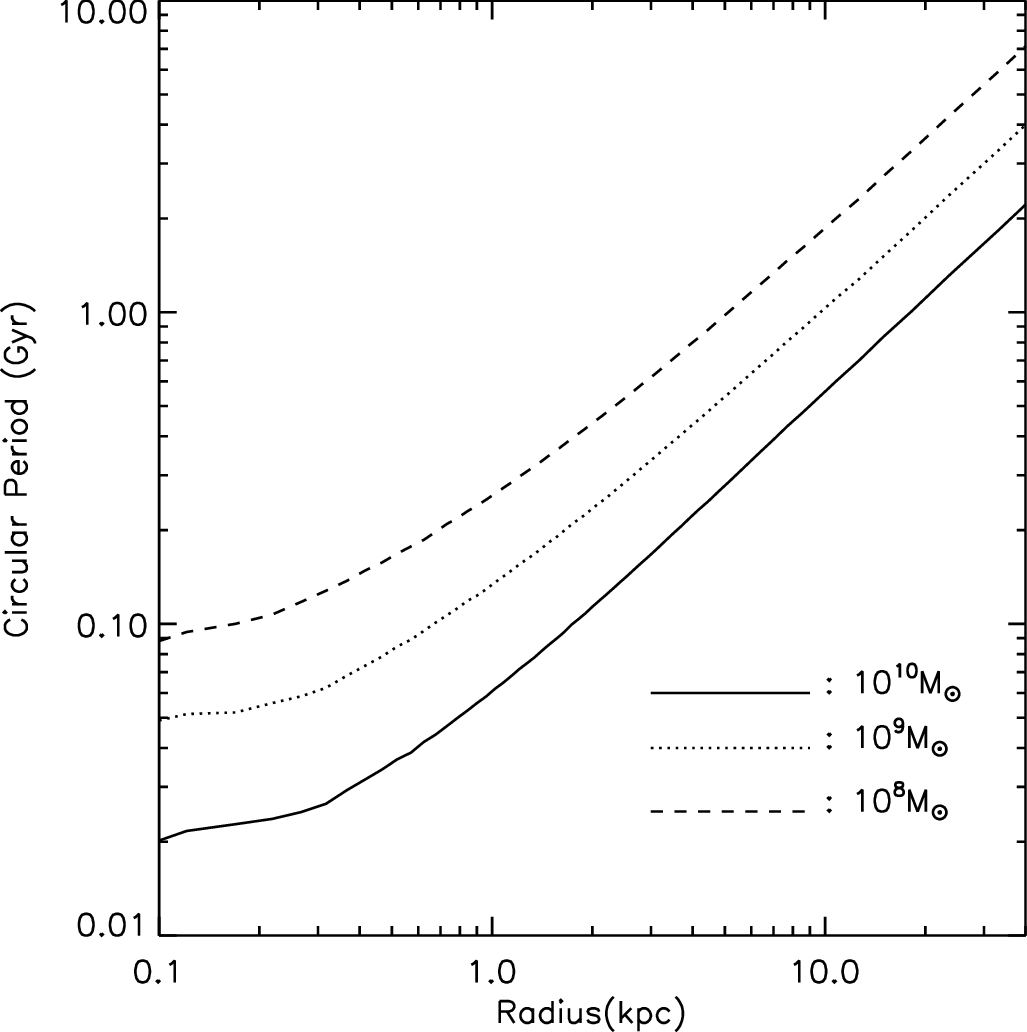}}\vskip 0.5cm
\makeatletter\def\@captype{figure}\makeatother
\caption{The period of 1:1 resonant circular orbits on the $x-y$ plane as a function of radius. The solid, dotted, and dashed lines are for models with mass of $10^{10}\Msun$, $10^{9}\Msun$ and $10^{8}\Msun$, respectively.}\label{period}
\end{center}
\end{figure}

In our simulations, we have $\mathit{N}=8\times 10^5$ particles for each model, and choose a grid for the numerical integration of Eq.~\ref{mond} with
$64\times 32 \times 64$ cells in the spherical coordinates $(r,
\theta ,\psi)$, where the radial grids are defined by $r_i=2.0\tan
[(i+0.5)0.5\pi /(256+1)]$kpc. 
The density is obtained via a
quadratic particle-mesh interpolation and the time integration is
performed by the classical two order leap-frog scheme. As our time unit
for all subsequent plots we use the following definition (cf. Wang et
al. 2008) \beq\label{time} T_{\rm simu}=\left( {GM\over
a^3}\right)^{-1/2}=4.7\times 10^6 yr\left({M\over
10^{10}\Msun}\right)^{-1/2}\left({a\over 1kpc}\right)^{3/2} . \eeq
which represents the Newtonian (or Keplerian) dynamical time at the radius of $r=a$ without the factor of $2\pi$. We remind the reader that the parameter $T_{\rm simu}$ is neither the dynamical time nor the orbital time in general MOND simulations. The orbital time in our MONDian systems is defined as the period of the 1:1 resonant orbit in the $x-y$ plane (Wang et al. 2008).
Fig.~\ref{period} shows the periods of circular orbits at different radii for the three models presented here. Fig.~\ref{period} as well as Eq.~\ref{time} imply that the MONDian dynamical time at the radius of 1 kpc is about $7 T_{\rm simu}$, $5T_{\rm simu}$ and $3.5T_{\rm simu}$ for the three models whose masses are $10^{10}\Msun$, $10^9\Msun$ and $10^{8}\Msun$. We further like to note that the internal time step used by the code NMODY to integrate the equations-of-motion is ${0.3\over \sqrt{\max |\nabla \cdot \bf g|}}$, where the factor $0.3$ is a typical number used in N-body simulations, and $\nabla \cdot \bf {g = 4\pi G \rho_{eff}}$, where $\rho_{eff} $ is the effective dynamical density of the system, i.e. the sum of the baryon and (phantom) dark matter density in the Newtonian force law to produce the gravity or potential of baryons in MOND (see We et al. 2008).  The time steps here are determined by the maximum values of $\nabla \cdot \bf {g} $, which means the densest dynamical region of the models, where gravity changes most sharply. Note that all particles share a common time step that typically is $0.005\sim0.03 T_{simu}$.

\begin{figure}{}
\begin{center}
\resizebox{11cm}{!}{\includegraphics{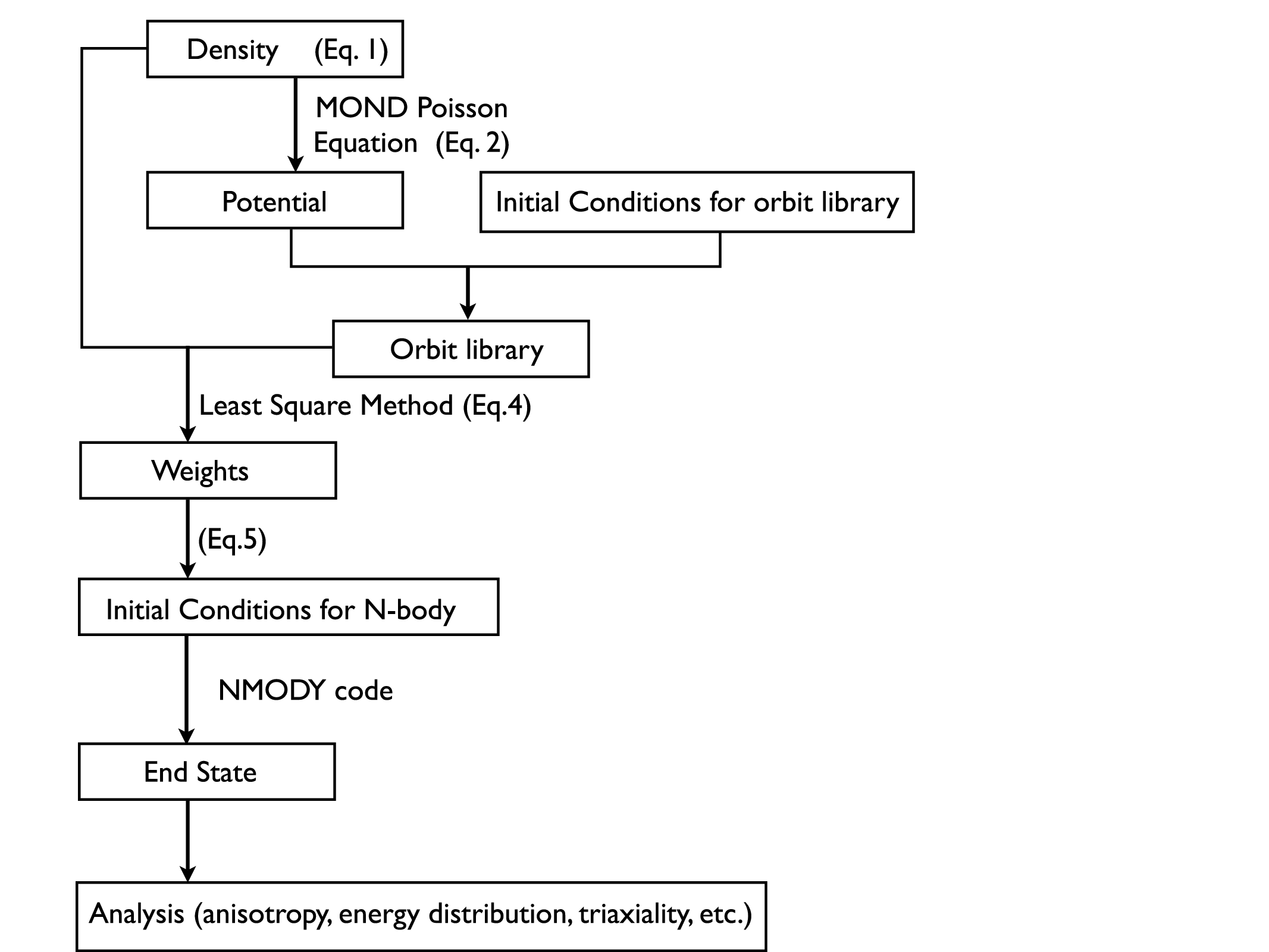}}
\makeatletter\def\@captype{figure}\makeatother
\caption{Flowchart of the simulations. }\label{flowchart}
\end{center}
\end{figure}

A flowchart of the technical steps involved in the process prior to the analysis stage can be viewed in Fig.~\ref{flowchart}. This figure summarizes the methodology of how to generate and evolve the $N$-body systems. In Table~\ref{TotalTimes} we present the total times each systems has been evolved for.

\begin{table}
\begin{center}\vskip 0.00cm
\caption{Total simulation times = $200\times T_{\rm simu}$.}
\begin{tabular}{llc}
\hline
 Model & N-body run duration $T$ & unit time $T_{simu}$\\
\hline
$10^{10}\Msun$ & 0.94 Gyrs & 4.7 Myrs\\
$10^{9}\Msun$  & 3.0 Gyrs &14.9 Myrs\\
$10^{8}\Msun$  & 9.4 Gyrs &47.0 Myrs\\
\hline
\end{tabular}
\label{TotalTimes}
\end{center}
\end{table}

\subsection{Virial Theorem}\label{virialratio}
\begin{figure}{}
\begin{center}
\resizebox{8.5cm}{!}{\includegraphics{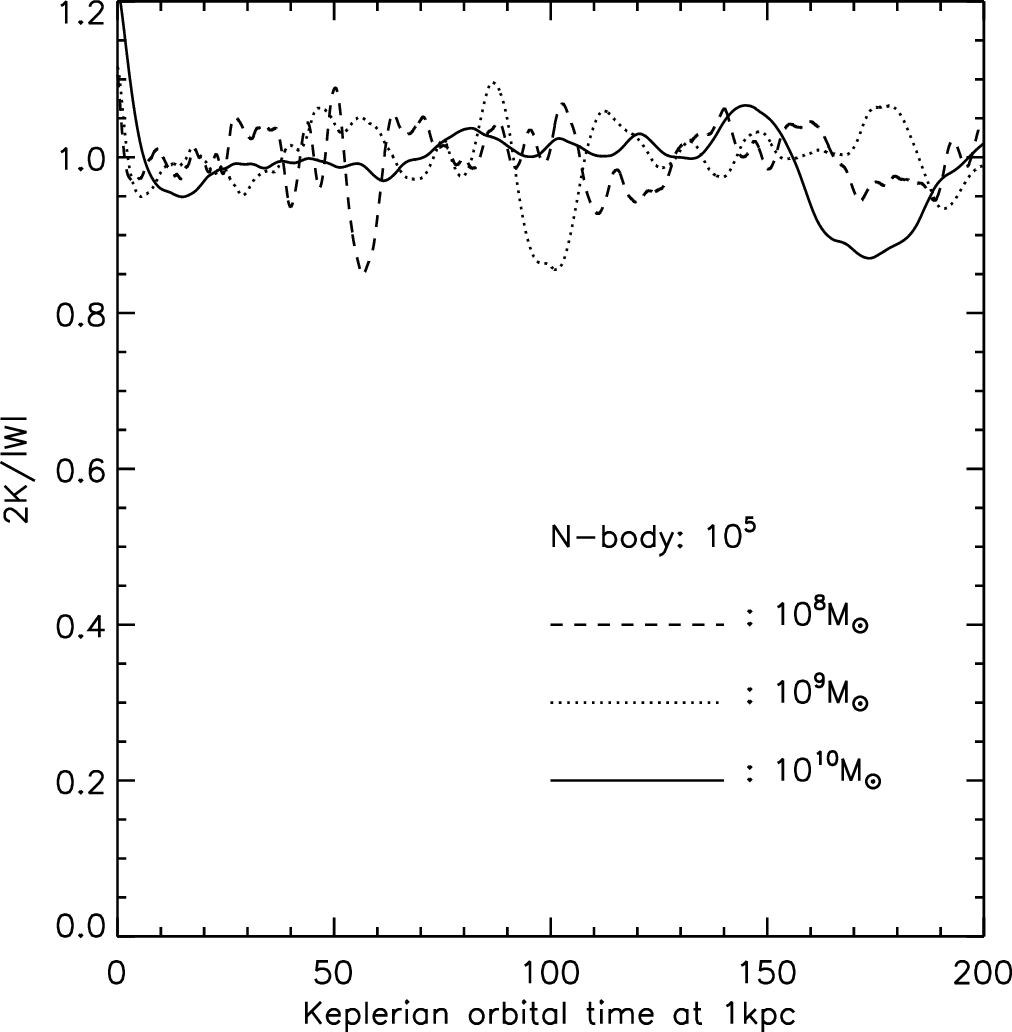}\includegraphics{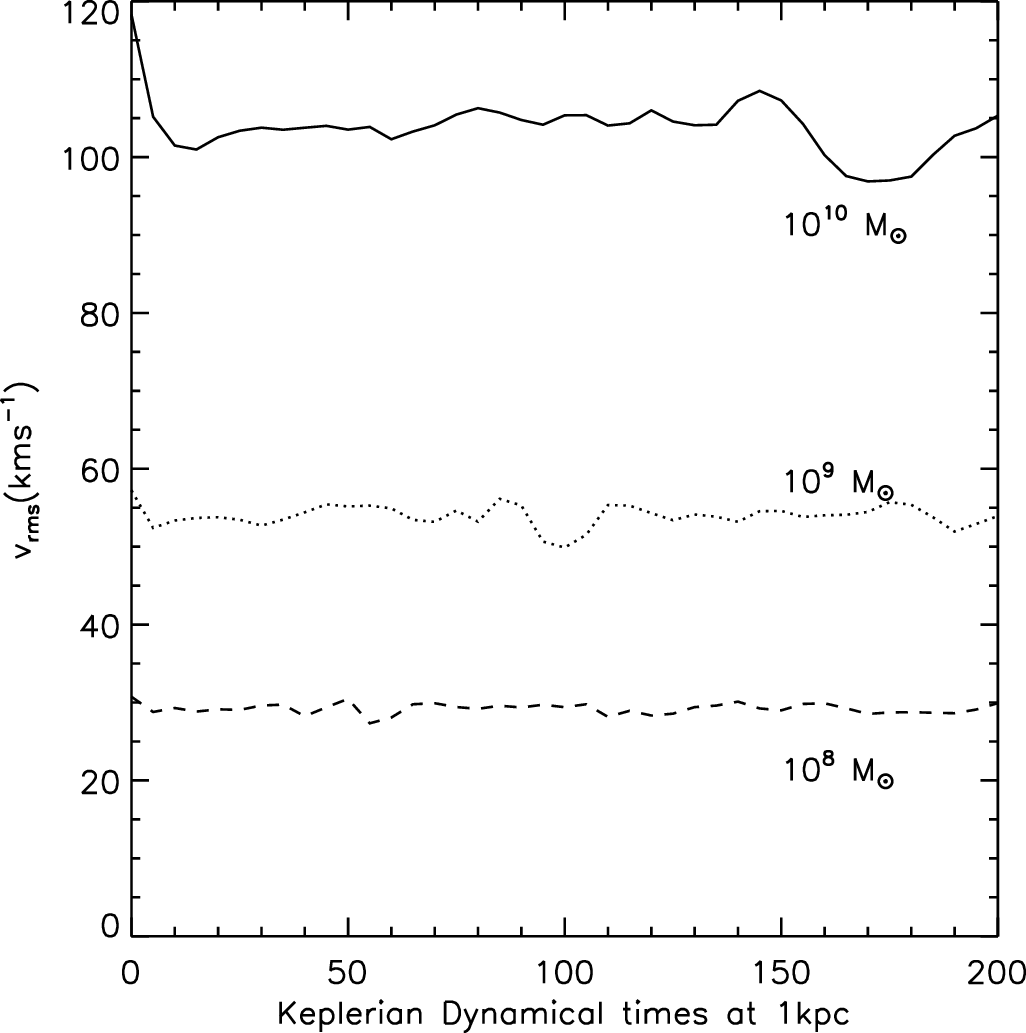}}
\makeatletter\def\@captype{figure}\makeatother
\caption{The evolution of $2K/|W|$ for all three systems. The solid, dotted, and dashed lines are for models with mass of $10^{10}\Msun$, $10^{9}\Msun$ and $10^{8}\Msun$, respectively. The evolution is shown for 200 Keplerian dynamical times (1.0 kpc).}\label{virial}
\end{center}
\end{figure}

The scalar Virial theorem, $W+2K=0$, is valid for systems in
equilibrium, where $W$ is the Clausius integral, \beq\label{Clausius} W=\int \rho\vec
x\cdot \nabla \Phi d^3x ,\eeq and $K$ is the kinetic energy of the
system (Binney \& Tremaine 1987). In the left panel of Fig.~\ref{virial},  we show that the evolution of $-2K/W$ for all
models is always about unity, as expected for an
equilibrium system. We though note that during the first circa five Keplerian
dynamical times (1.0 kpc) all systems are moving from a
quasi-equilibrium state with $-2K/W\approx 1.1 - 1.2$ to $-2K/W =
1.0 \pm 0.1$ afterwards (marginally oscillating about unity). This figure demonstrates that our $N$-body ICs
start off in quasi-equilibrium and after approximately five
Keplerian dynamical times (1.0 kpc) can be considered fully relaxed.
These 'hot' $N$-body ICs could be due to a number of reasons including the resolution of the simulation and
chaotic orbits, respectively. Regarding the latter, we need to mention that we compute the orbit library for 100
orbital times, and the time integration may not be
long enough to ensure a relaxation of those chaotic orbits; particles coming from the chaotic orbits could lead to higher pressure 'overheating' the
system.

In Fig.~\ref{virial}, we plot the velocity dispersion $v_{\rm rms}$ for all systems as a function of the simulation time unit. The plot indicates that each $v_{\rm rms}$ decreases by about 10\% during the relaxation process and stays constant afterwards (with tiny variations though).\footnote{There are typos in Wang et al. (2008) about the total mass of models and $v_{rms}$.}


Note that (as inferred from the right panel of Fig.~\ref{virial}) the kinetic energy of the systems decrease nearly one quarter for the maximal evolved case after the relaxation.\footnote{The kinetic energy $K$ is proportional to $v_{\rm rms}^2$.} However, the Virial theorem is still satisfied. That does \textit{not} mean the energy conservation law is broken: $W+K$ is not the total energy of a MONDian system, and it isnot conserved either. The total energy is still the conserved quantity but for a MONDian system it is given by (Bekenstein \& Milgrom 1984):
\beq\label{energy}
E=-L+K
\eeq
where $L$ is the Lagrangian of the MONDian system, defined by \beq L=\int d^3 r \left\{ \rho\Phi+{1\over 8\pi G}a_0^2\mathcal{F}\left[{(\nabla \Phi)^2\over a_0^2} \right]\right\},\eeq and $\mathcal{F}(x^2)$ is an arbitrary function with $\mu(x)=\mathcal{F}'(x^2)$. For an isolated system in MOND, the potential is logarithmic thus the potential energy is infinite. Therefore, the only meaningful quantity is the difference in energies between different systems (Bekenstein \& Milgrom, 1984; Nipoti et al. 2007). 
However, the evident evolution of $W+K$ at the very beginning (i.e. the first 5 Keplerian dynamical times (1.0 kpc)) shows that the N-body ICs are not accurately in equilibrium, and hence referred to as quasi-equilibrium. 


\subsection{Energy Distribution} \label{energydistributions}
One of the characteristic quantities to describe relaxation
processes is the so-called differential energy distribution, i.e. the quotient of 
mass $dM$ over the energy band interval $[E,E+dE]$ (Binney \& Tremaine
1987). The energy of a unit mass element is $E={1\over
2}v^2+\phi(\vec x)$, where $\phi$ is logarithmically infinite in
MOND and hence all particles are bound. But since the absolute value
of potential energies is meaningless, we can define the zero point
as the last point of the radial grid. Hence, there are positive
relative energies for part of the particles though all of them are
bound to the system.

\begin{figure}{}
\begin{center}
\resizebox{8.cm}{!}{\includegraphics{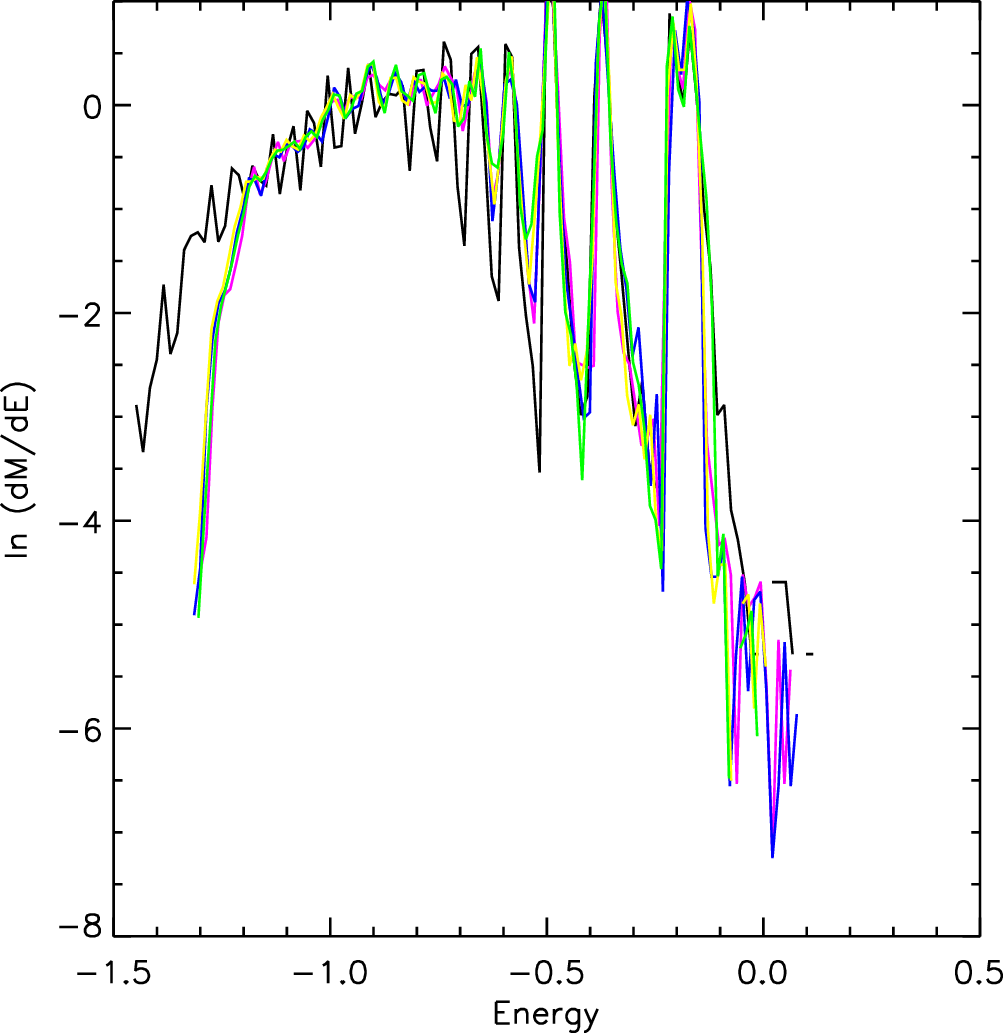}\includegraphics{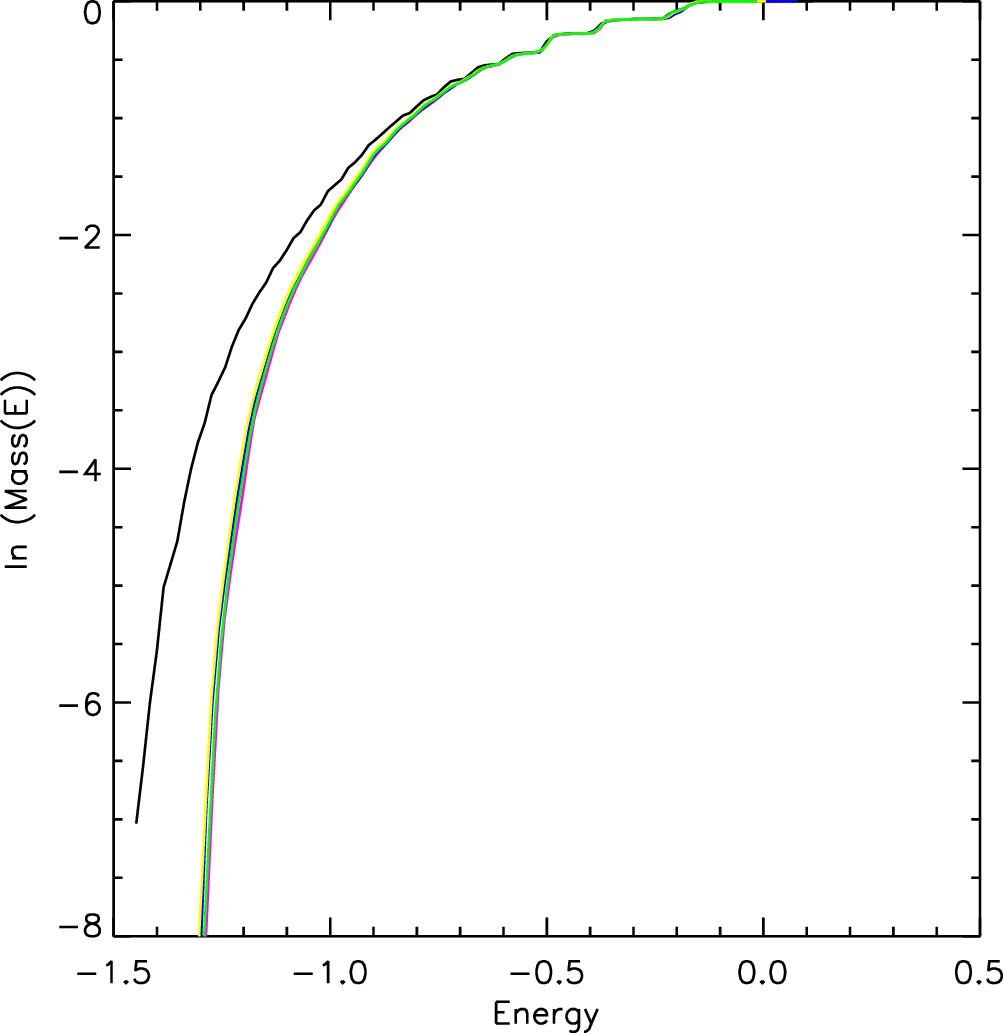}}\vskip 0.5cm
\resizebox{8.cm}{!}{\includegraphics{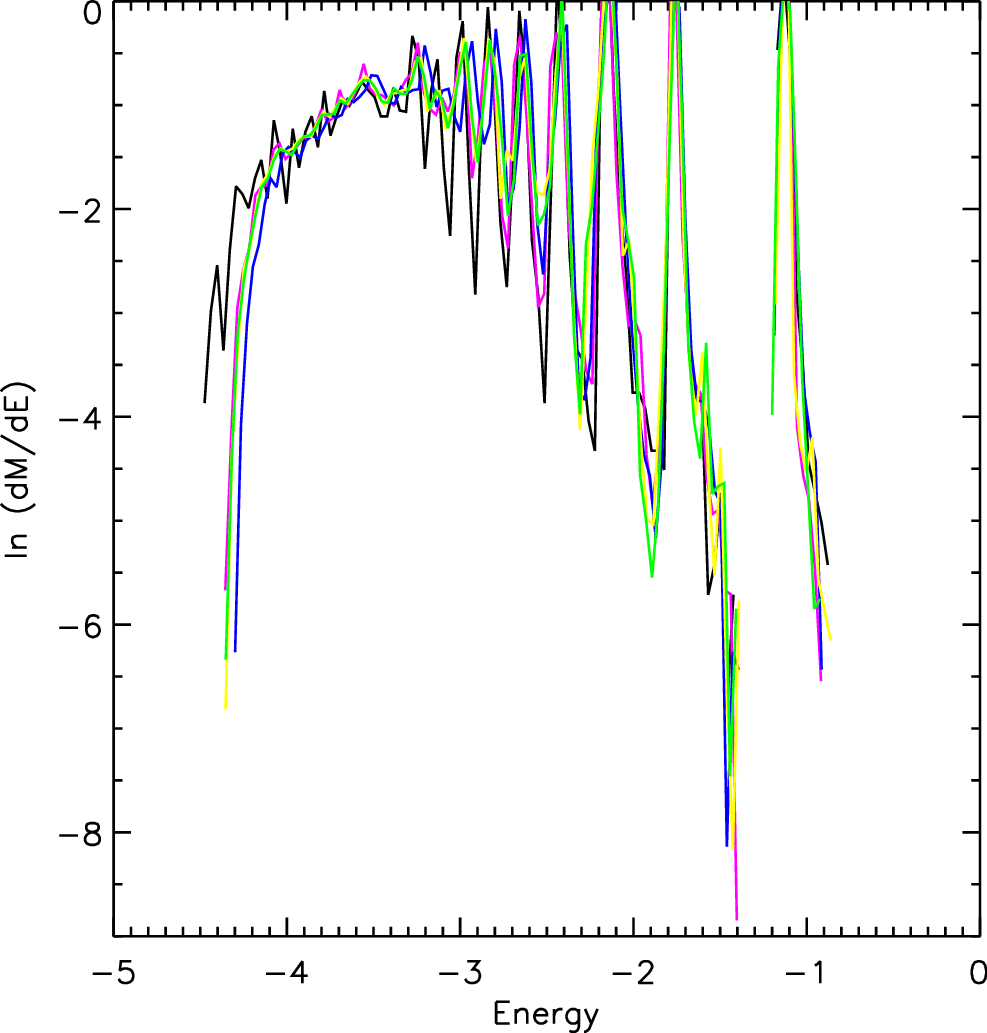}\includegraphics{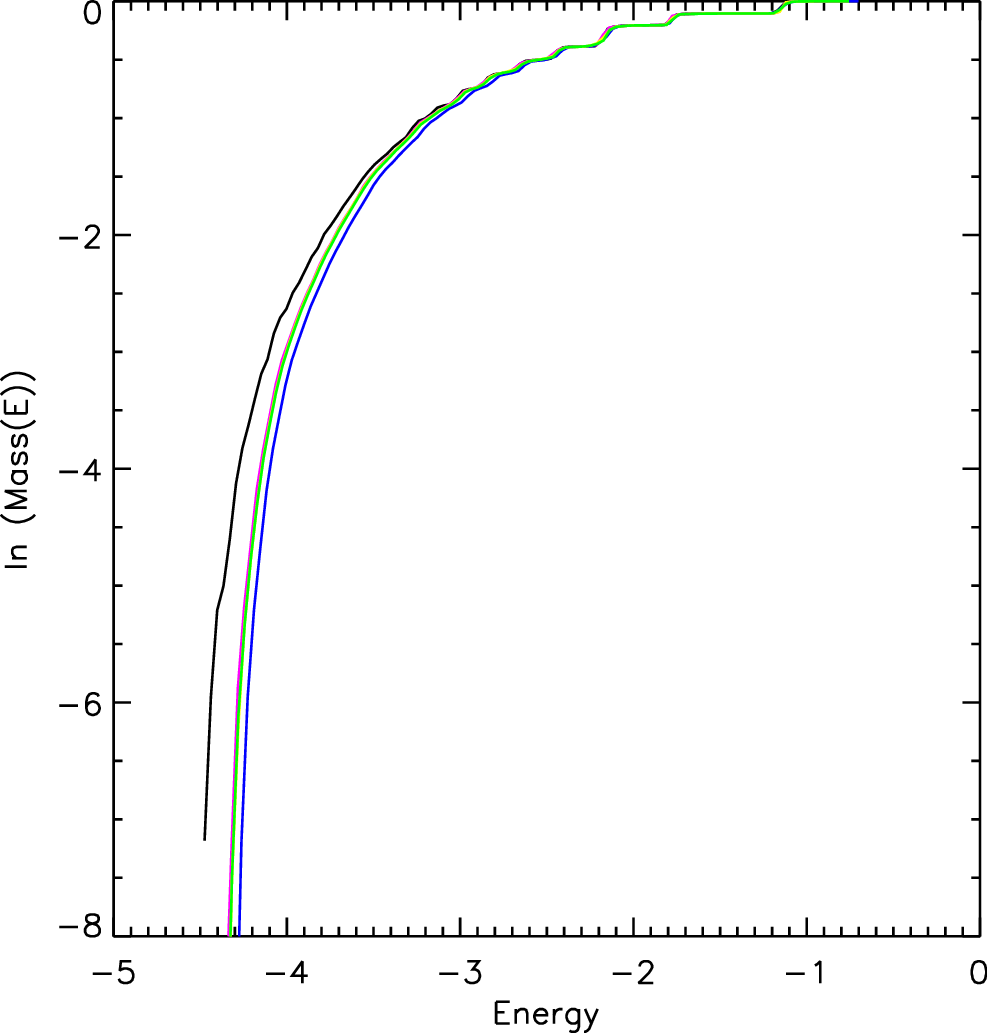}}\vskip 0.5cm
\resizebox{8.cm}{!}{\includegraphics{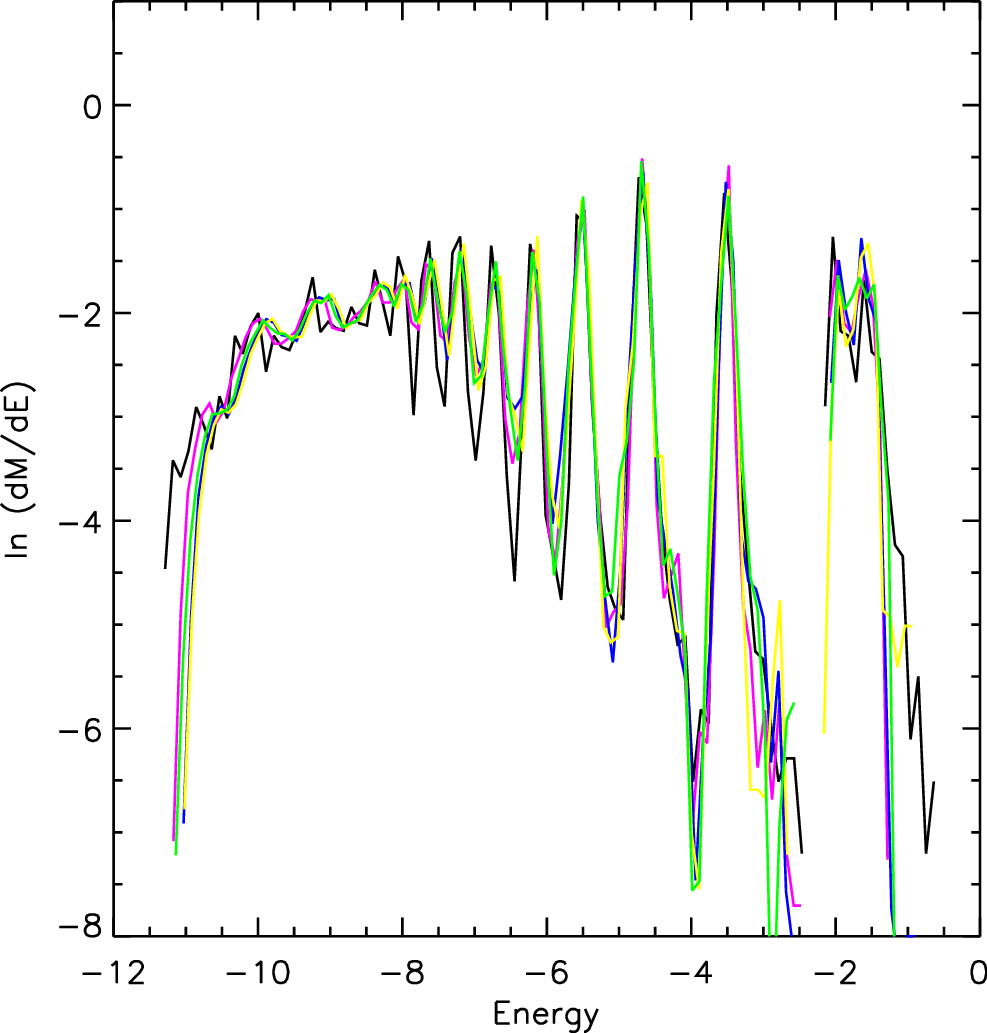}\includegraphics{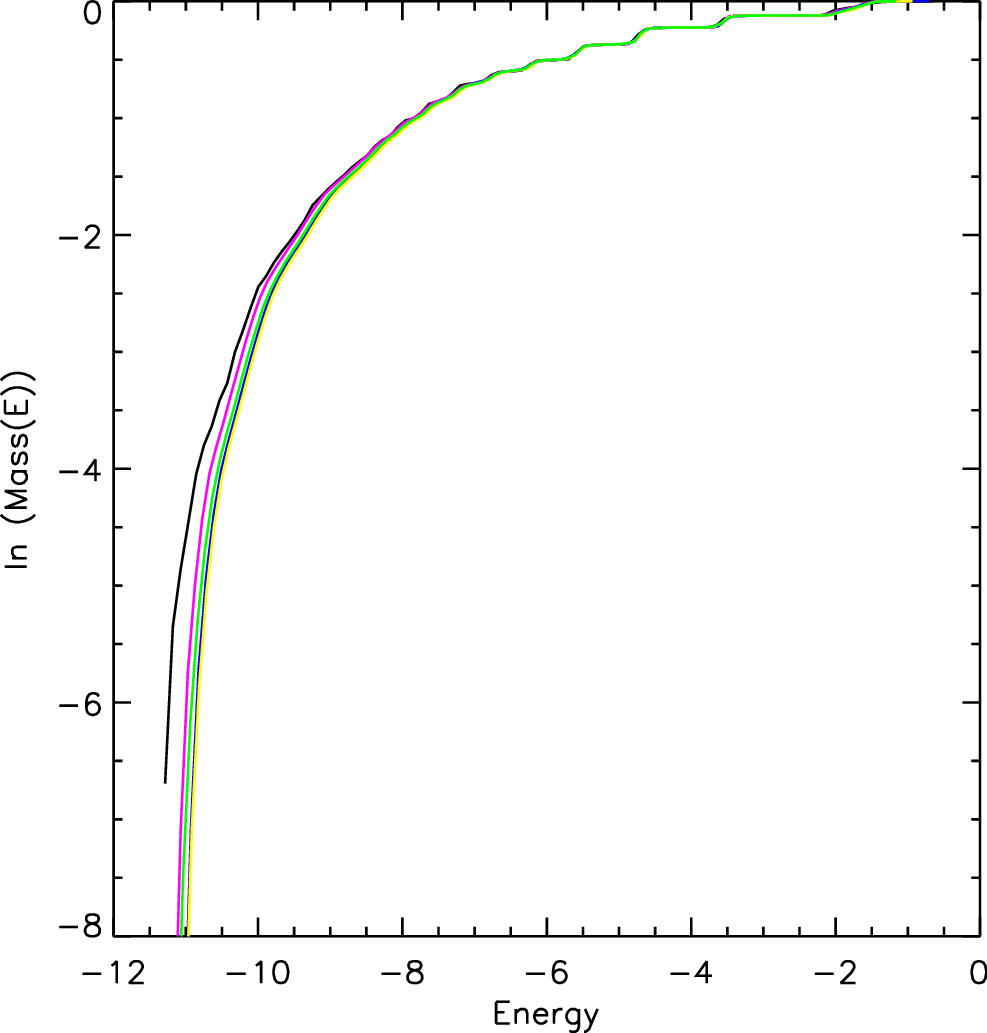}}\vskip 0.5cm
\makeatletter\def\@captype{figure}\makeatother
\caption{\textbf{Left panels}: Evolution of the differential energy distribution ${dM\over dE}$. The panels (from upper to lower) correspond to our models with total mass of $10^{10}\Msun$, $10^{9}\Msun$ and $10^{8}\Msun$, respectively. \textbf{Right panels}: The accumulation of energy distribution. The black lines denote the ICs, and the violet, blue, yellow and green lines show the differential energy after 50, 100, 150, 200 Keplerian dynamical times (1.0 kpc). Both ${dM \over dE}$ and the Energy are given in units where G=1 and M=1.
}\label{dmde}
\end{center}
\end{figure}

The left panels of Fig.~\ref{dmde} show the evolution of ${dM\over dE}$ over 200 Keplerian dynamical times (1.0 kpc) for all three models (upper to lower) and we find that all distributions are rather similar. And the most pronounced evolution of the energy distribution is at the low-$E$ end, where particles are most strongly bound to the system. All of the differential energy distributions have 19 peaks, as can be seen in left panels of Fig. ~\ref{dmde} That is due to the energy definition of the Schwarzschild technique outlined in \S \ref{schw}: Inside every sector, the energy (kinetic plus potential energy) is a constant, while the outer shell is the zero-velocity surface of this sector. The adjacent two sectors have energy jumps at the shell. Therefore there are 19 'quantized energy levels' for our models, and for each model there are no mass distributions outside these 19 constant 'energy levels' and hence they appear as 'valleys' in the left panels of Fig.~\ref{dmde}. That explains why the curves appear noisy.
We note that after the initial relaxation of about five Keplerian dynamical times (1.0 kpc)\footnote{Even though we do not show the curves for 5 Keplerian dynamical times (1.0 kpc) we acknowledge that the drop happens during that initial relaxation phase. Here we care about the long-term evolution within 200 Keplerian dynamical times (1.0 kpc) and hence decided to rather focus on the late evolution of the systems.} the low-$E$ end of the distribution becomes devoid of particles, i.e., particles are leaving the central regions where the potential well is deepest. This actually hints at a possible flattening of the initially present density cusp $\gamma=1$! We return to this issue later in sub-section~\ref{massdistributions}.

Comparing the three left panels in Fig.~\ref{dmde}, we observe that the system in the mild MOND regime (i.e. the model with a mass of $1\times 10^{10}\Msun$: upper-left panel) has the most significant evolution, whereas the model in deep MOND evolves least (lower-left panel). We therefore conclude that our ICs are most stable for the deep MOND regime.

As seen in the right panels of Figure \ref{dmde}, the accumulation of energy distribution clearly confirms the previous conclusion from the differential distributions. After the relaxation, the mass in the inner region escapes to the outer, while the outer part is nearly unchanged. The mass distribution obviously does not evolve after relaxation. 

\subsection{Kinetics} \label{kinetics}
To further check upon the stability of our systems, we calculate the radial velocity dispersion profiles $\sigma_r(r)$ as well as the anisotropy parameter \beq\beta(r)\equiv 1-{\sigma_\theta^2+\sigma_\psi^2 \over 2\sigma_r^2}.\eeq Here $r$ is the spheroidal radius, the same as in Eq.~\ref{den}, and $\sigma_\theta$, $\sigma_\psi$ are the tangential and azimuthal velocity dispersions.

\begin{figure}{}
\begin{center}
\resizebox{8.cm}{!}{\includegraphics{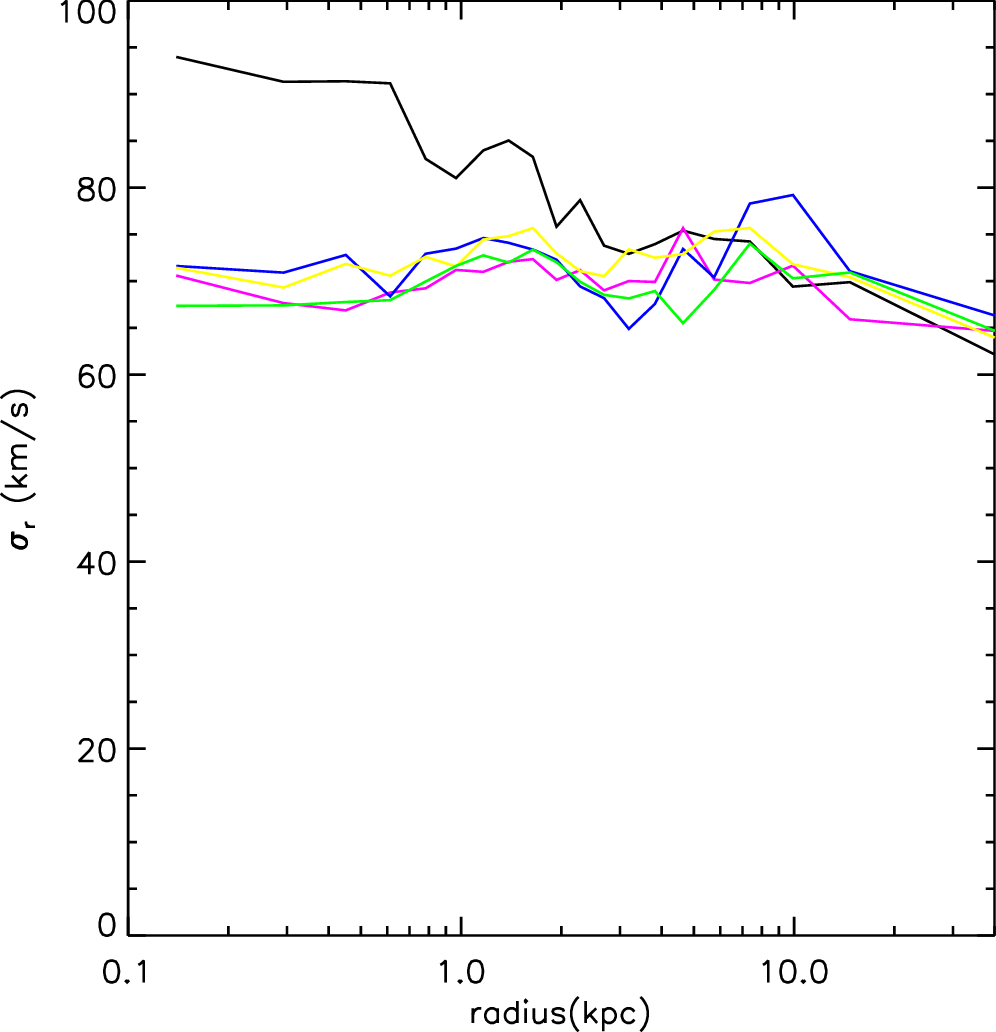}\includegraphics{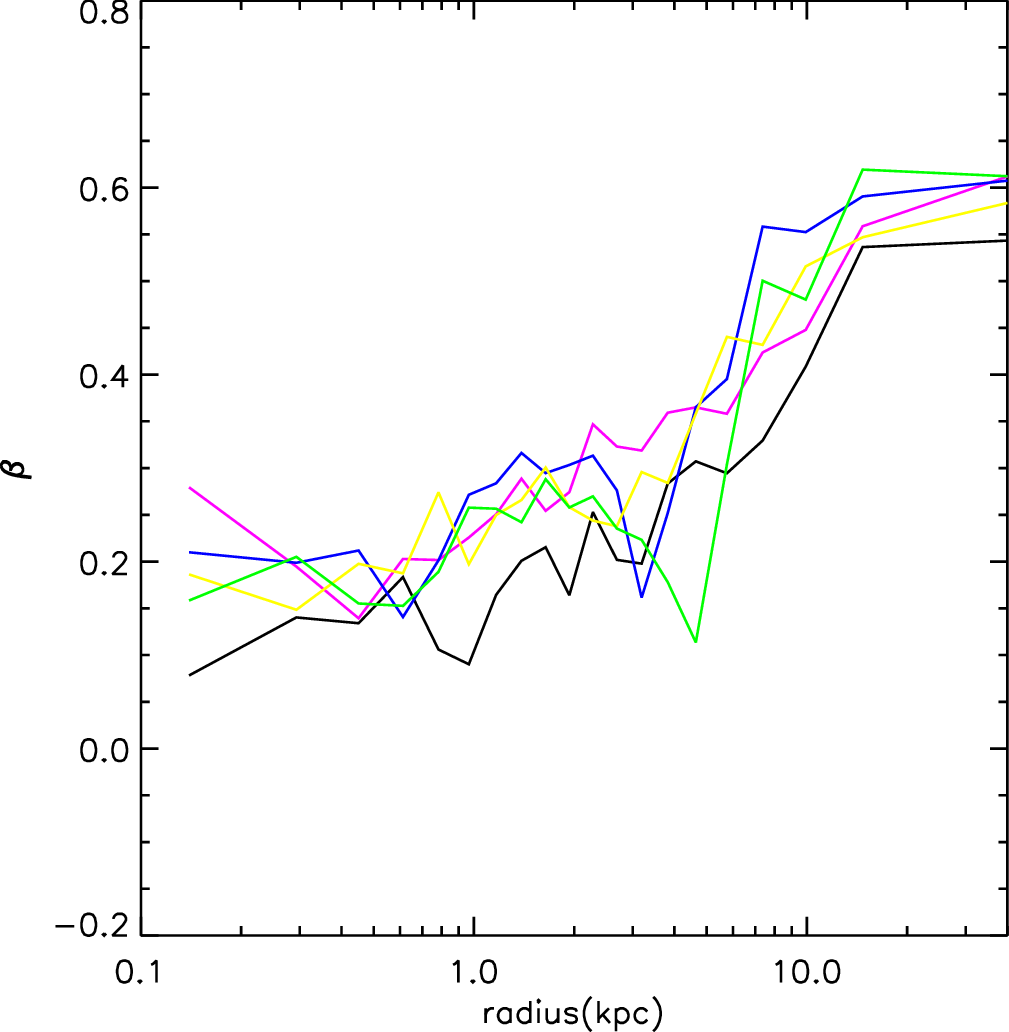}}\vskip 0.5cm
\resizebox{8.cm}{!}{\includegraphics{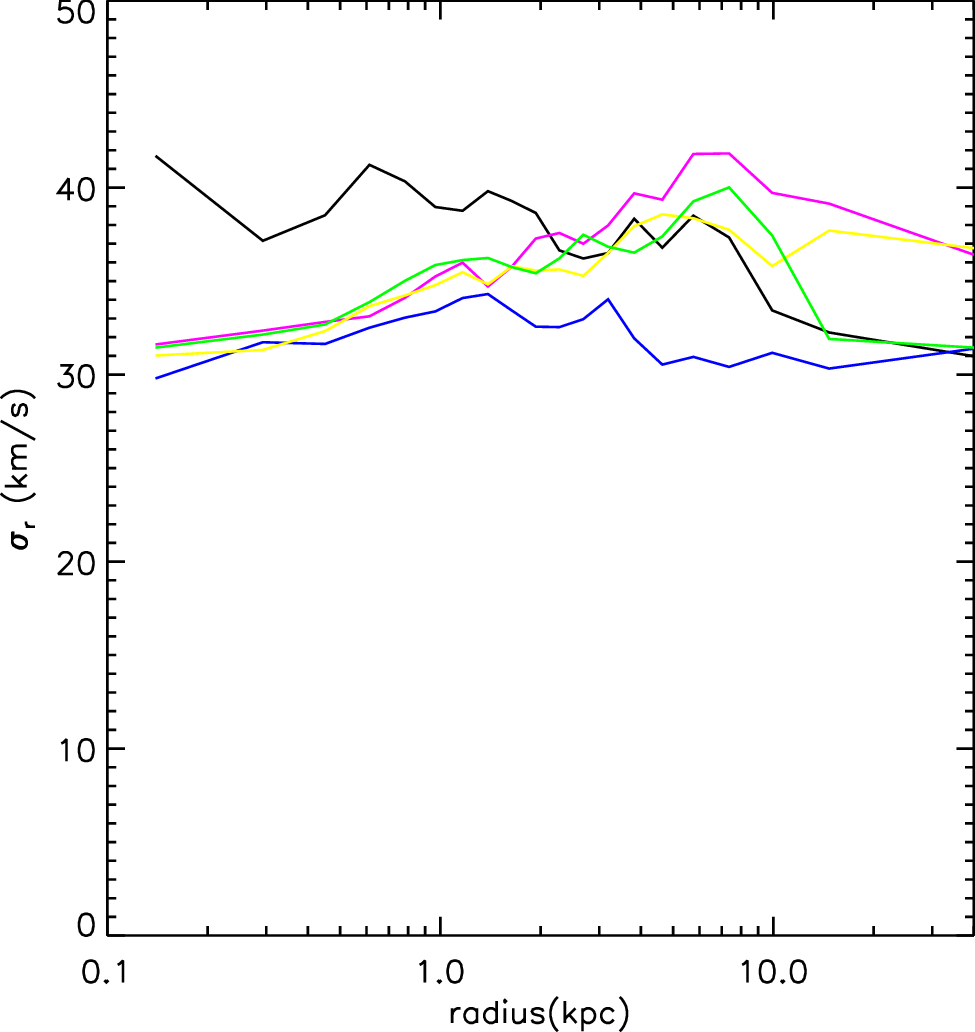}\includegraphics{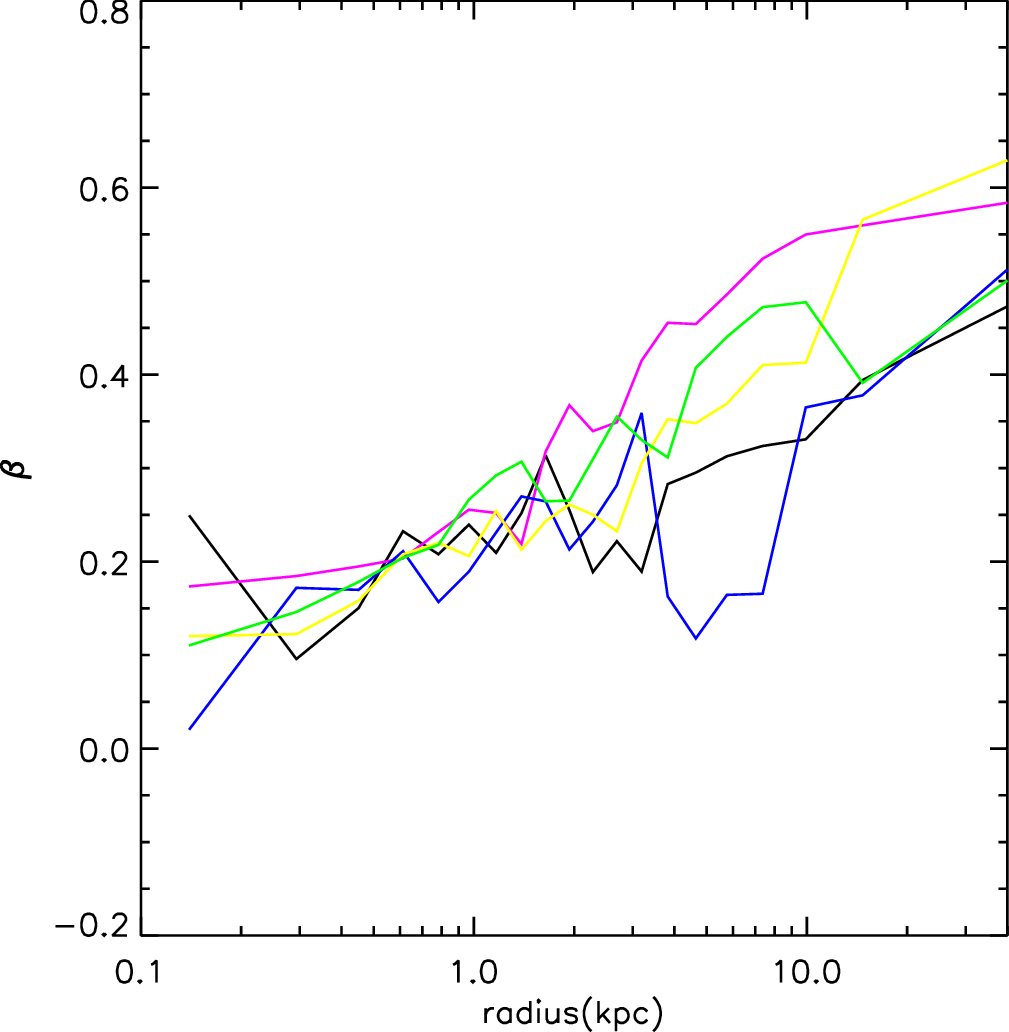}}\vskip 0.5cm
\resizebox{8.cm}{!}{\includegraphics{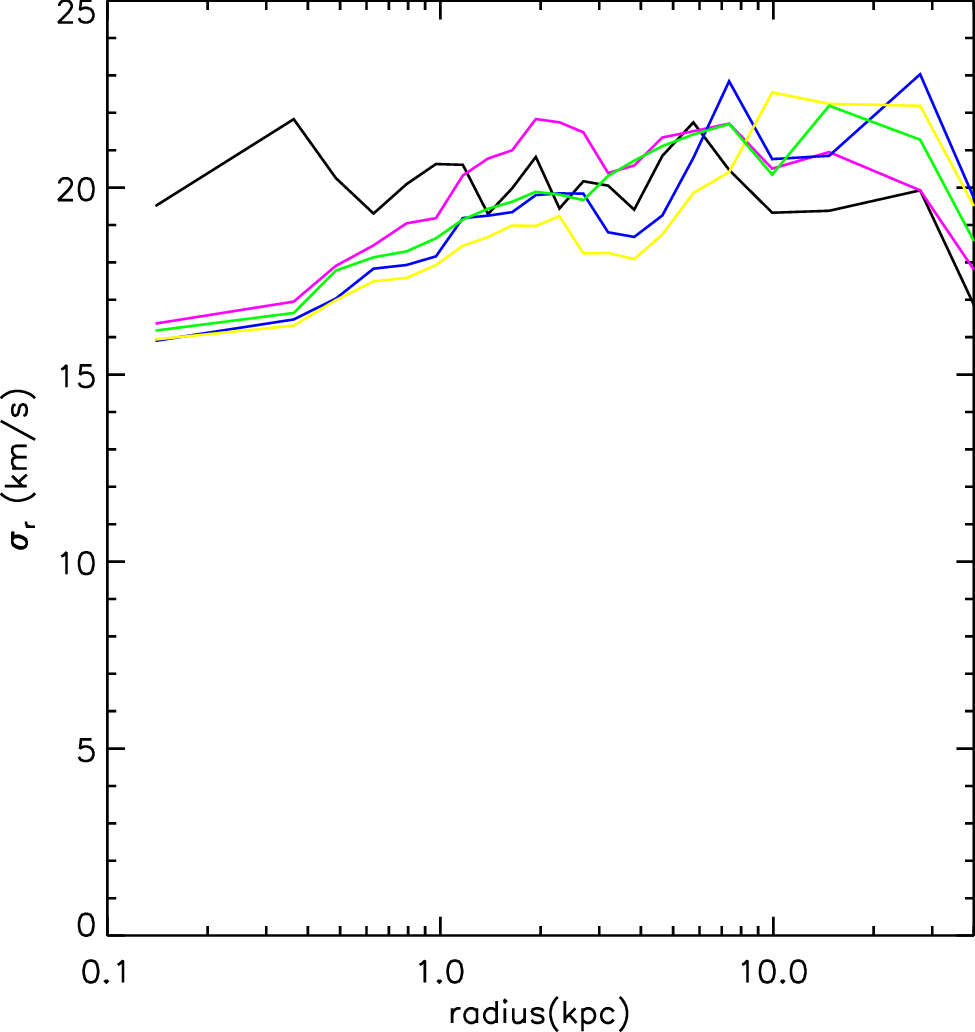}\includegraphics{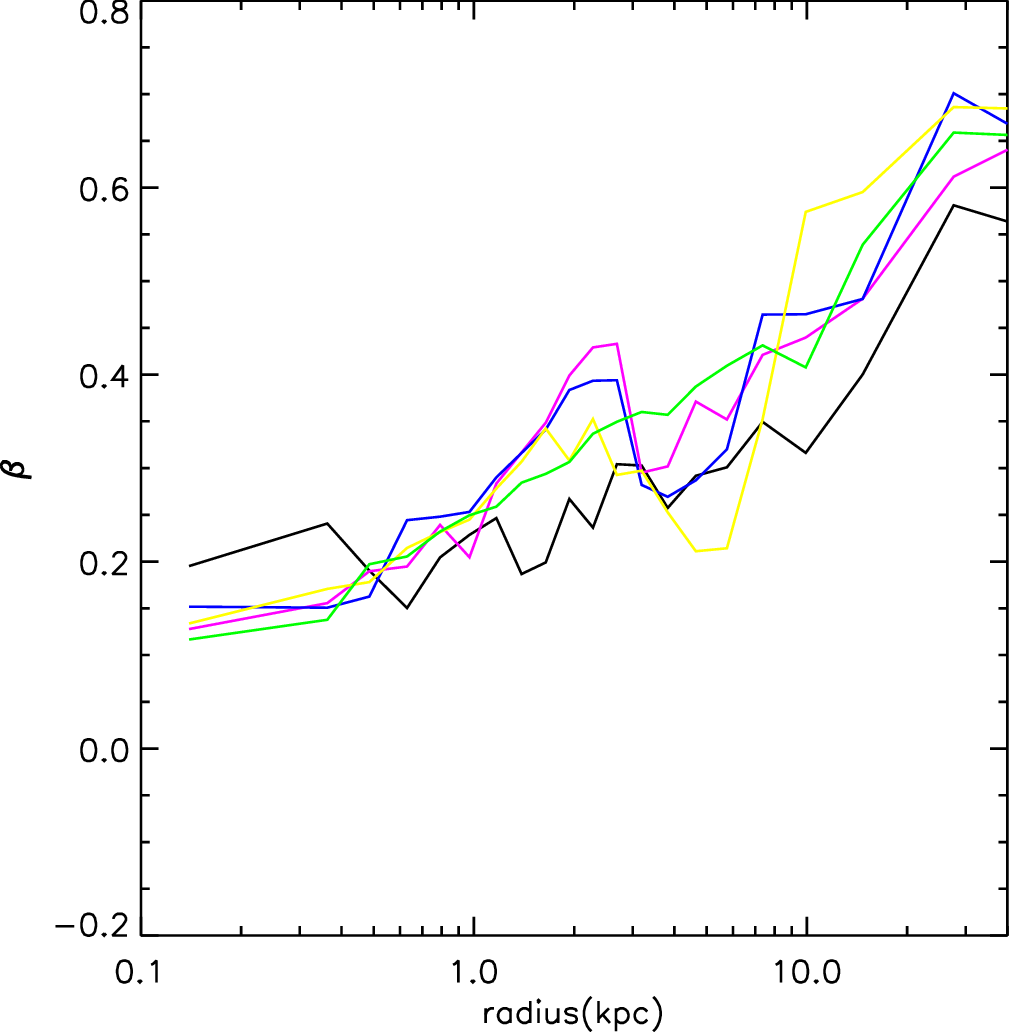}}\vskip 0.5cm
\makeatletter\def\@captype{figure}\makeatother
\caption{\textbf{Left panels:} Evolution of the radial velocity dispersion $\sigma_r(r)$. \textbf{Right panels:} Evolution of the velocity dispersion anisotropy $\beta(r)$. The \textbf{upper}, \textbf{middle} and \textbf{lower panels} are corresponding to models of $M=1.0\times10^{10}\Msun$, $M=1.0\times10^{9}\Msun$ and $M=1.0\times10^{8}\Msun$. The ordering of the panels corresponds to Fig.~\ref{dmde} as does the colouring of the lines.}\label{sigmar}
\end{center}
\end{figure}

The results can be viewed in the left panels of Fig.~\ref{sigmar}. We find that, for all three models, $\sigma_r$ drops within the central 2~kpc during the relaxation at the beginning of the simulation, i.e. the first five Keplerian dynamical times (1.0 kpc) (though not shown for clarity). 
Afterwards, there is only very little evolution noticeable. The reduction of $\sigma_r$ in the core means that the ICs are too hot in the radial direction to sustain equilibrium. We also note that the drop is more pronounced the less MONDian the ICs are. In the mild MONDian model with a mass of $M=10^{10}\Msun$, (upper-most panel) the model obviously appears to be hotter inside than outside. This trend is weakened for the deep-MOND model with a total mass of  $M=10^8\Msun$ where the self-gravity of the system is much weaker than $a_0$. The slope of $\sigma_r(r)$ oscillates around a constant value of approximately 20~km/s. In the intermediate model with $M=10^{9}\Msun$ the slope is between the most massive and least massive ones.

However, after the relaxation, the slope of $\sigma_r(r)$ has the same behavior in all three models, radially cooling down towards the cores. The curves of velocity dispersion after relaxation look like the rotation curves of disc galaxies at the similar mass range (Milgrom \& Sanders 2007; Gentile 2008). 
 In the centres, the tangential velocity dispersion $\sigma_\theta^2+\sigma_\psi^2$ plays an important role for the Virial Theorem and keeps the cores in equilibrium. Therefore, the systems prefer more isotropic velocity dispersions in the cuspy centres. To confirm this, we also present the anisotropy parameter $\beta(r)$ in the right panels of Fig.~\ref{sigmar}. We do find the expected small values of $\beta$ in the centres as well as a radial increase of $\beta(r)$. Inside 1~kpc the $\beta$-profiles oscillate during the whole evolution while they remain stable in the outer parts. Furthermore, the anisotropy increases with radius in all three models out to about 25~kpc where it turns approximately constant, $\beta=0.6$, i.e., the velocities are distributed hyper-radially.

Nevertheless, within 2~kpc, there should be a more substantial redistribution of kinetic energies in the tangential direction after relaxation due to the evolution of $\sigma_r$ seen in the left panels of Fig.~\ref{sigmar}; otherwise, the systems lose quite a lot of kinetic energy in the core. Unfortunately, the evolution of $\beta$ in the central region, seen in right panels of Fig.~\ref{sigmar}, is not as large as expected. Thus, there are outflows of kinetic energy from the centres of the systems. The values of velocity dispersion and anisotropic parameters in the deep MOND regions of our systems fit pretty well with the analytical predictions of isothermal spheres by Milgrom (1984; 1994).


\subsection{Mass distributions}\label{massdistributions}
\begin{figure}{}
\begin{center}
\resizebox{8.5cm}{!}{\includegraphics{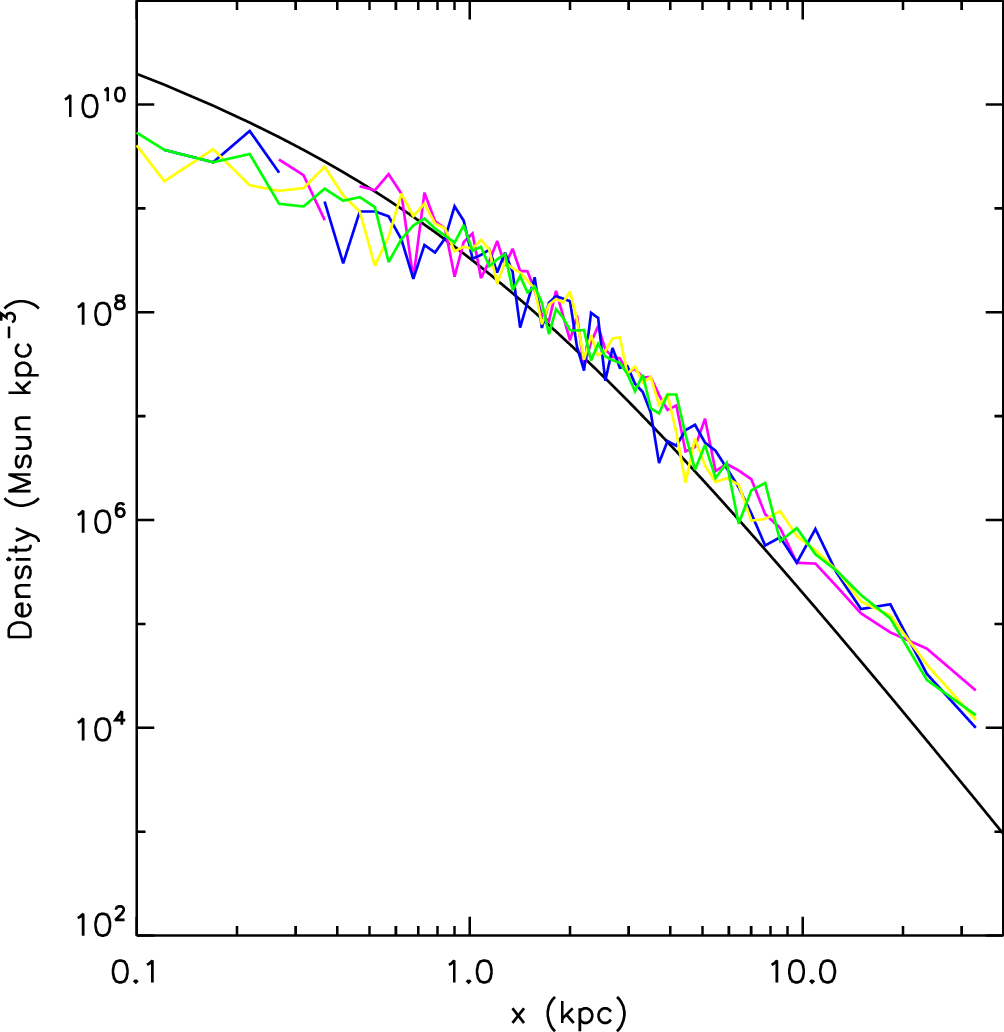}\includegraphics{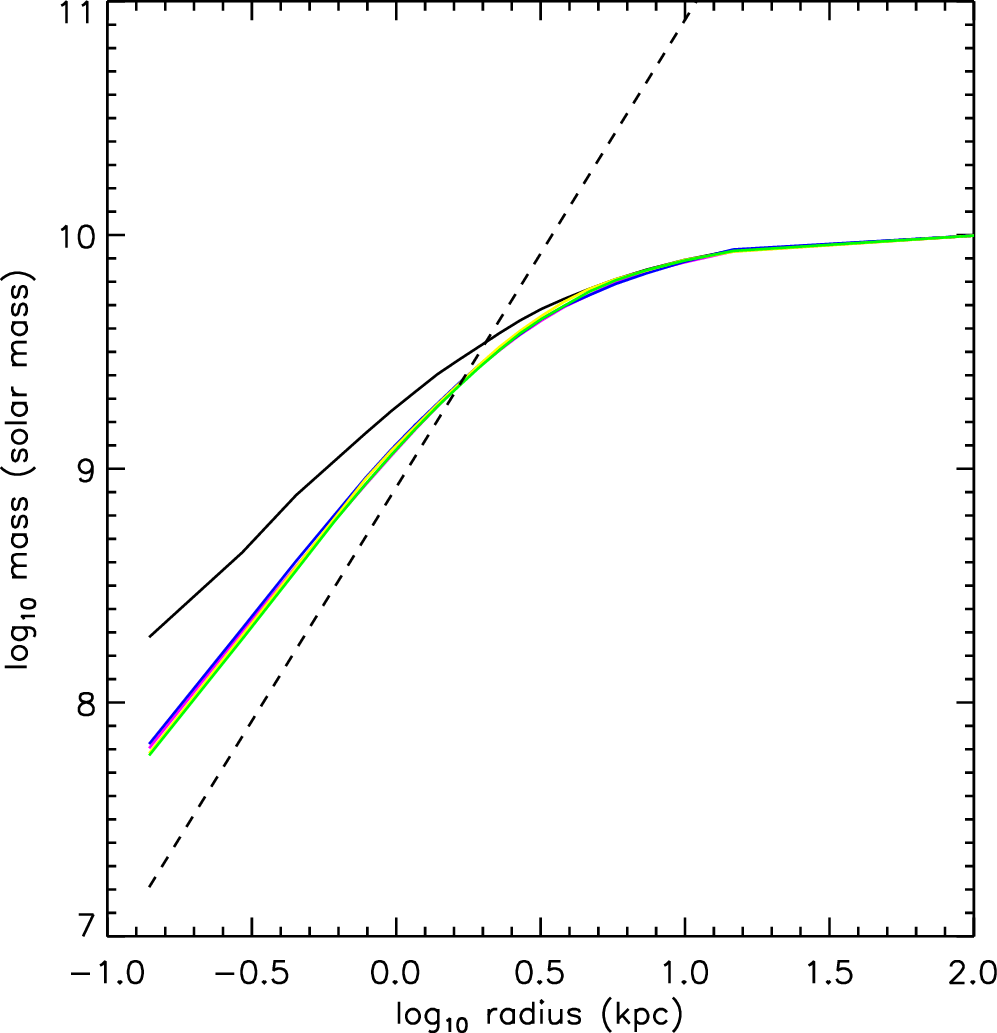}}
\resizebox{8.5cm}{!}{\includegraphics{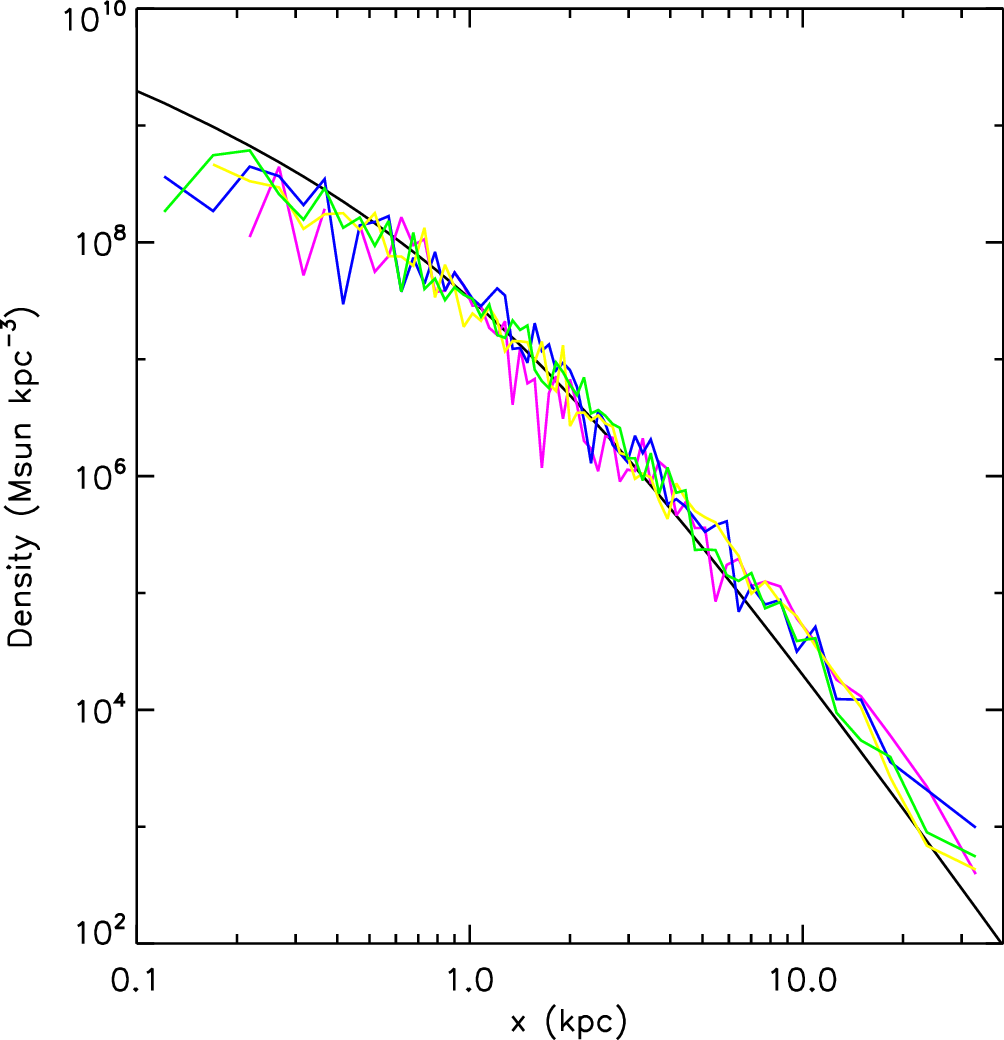}\includegraphics{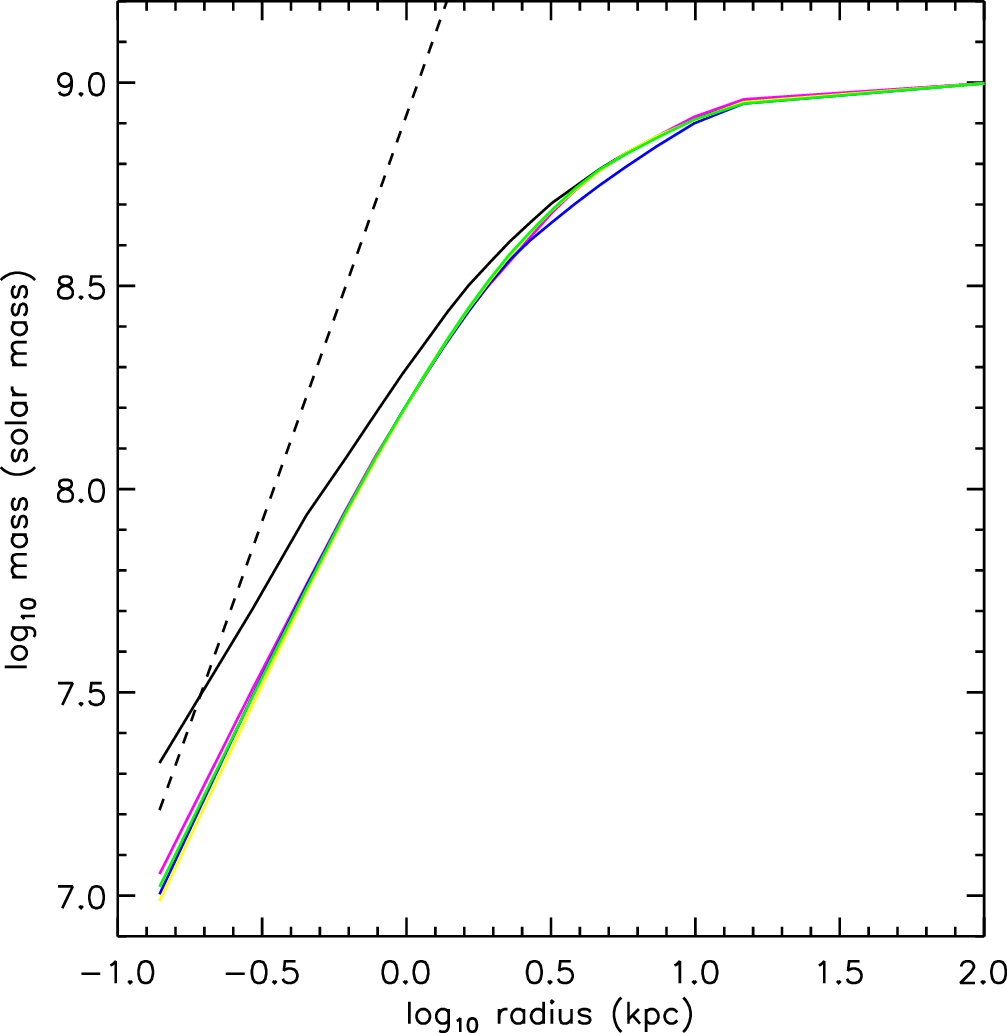}}
\resizebox{8.5cm}{!}{\includegraphics{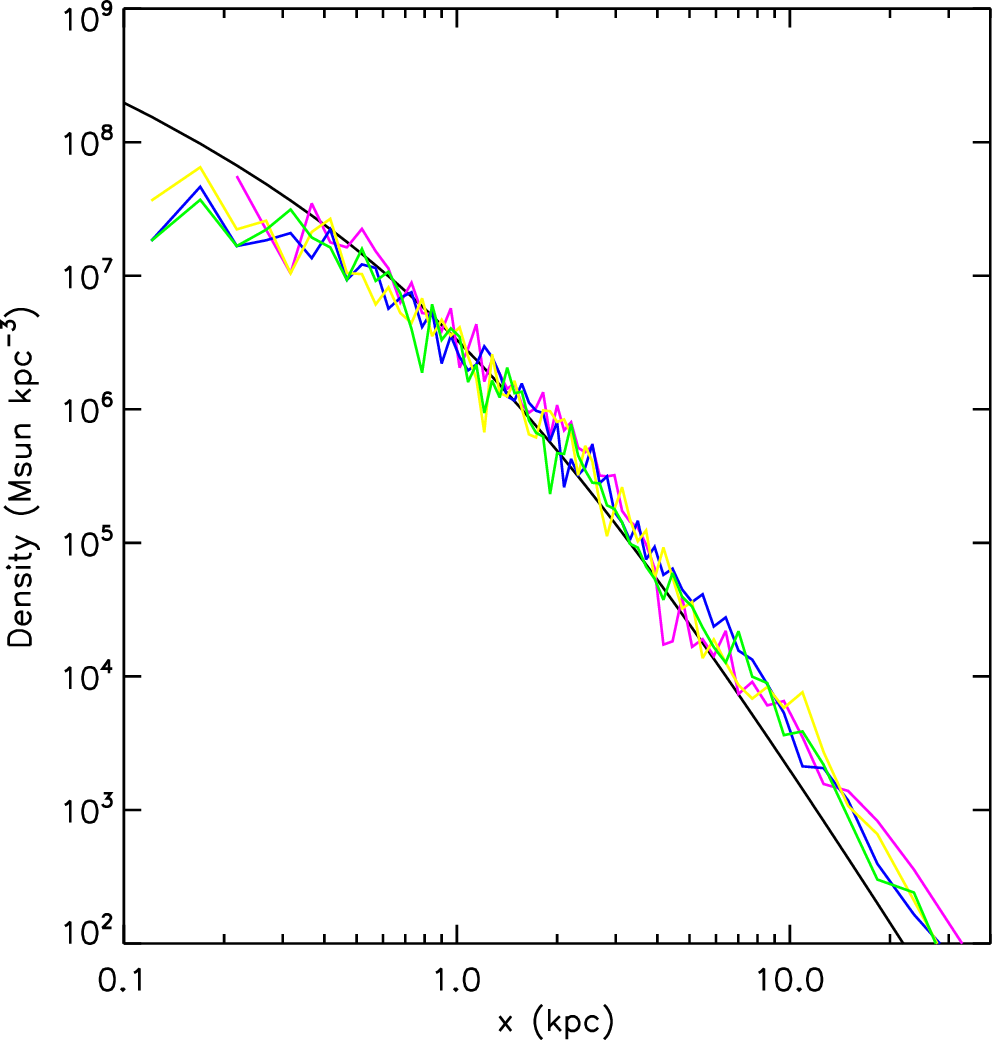}\includegraphics{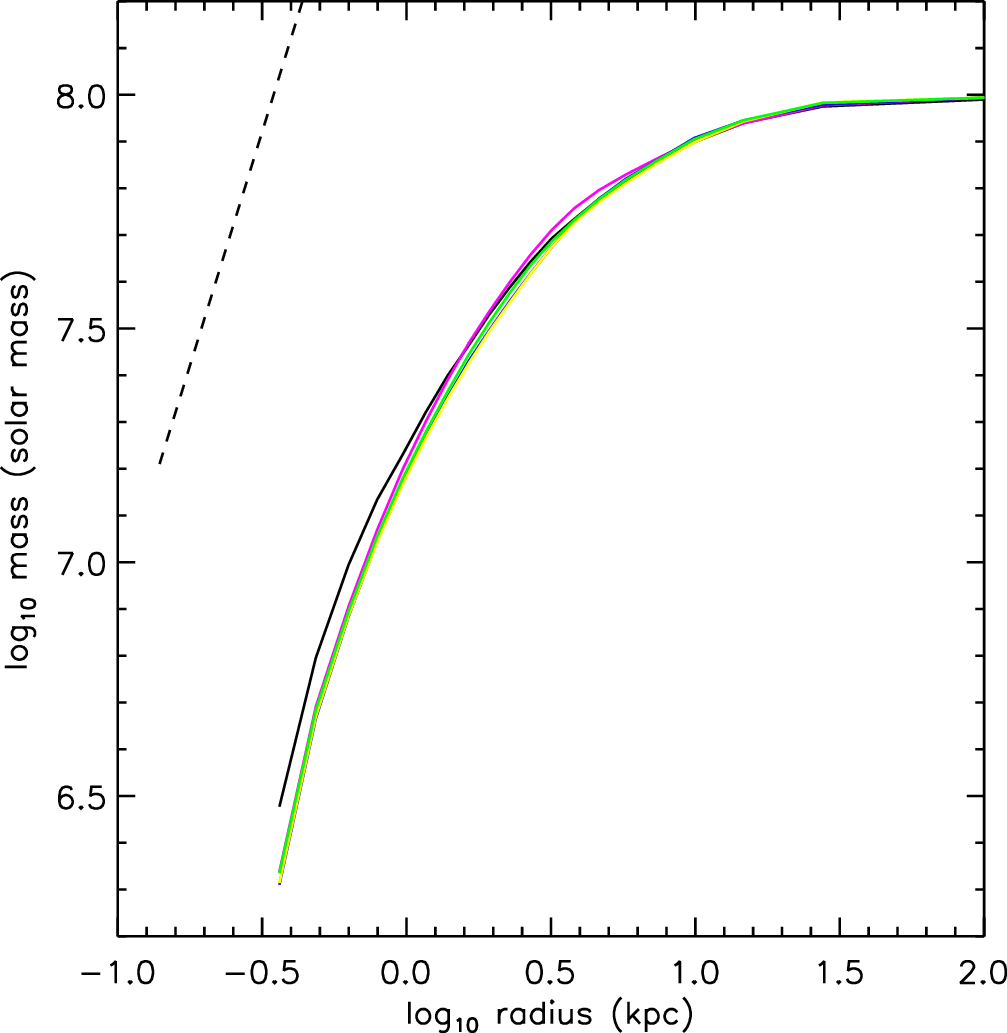}}
\vskip 0.5cm
\makeatletter\def\@captype{figure}\makeatother
\caption{The evolution of the mass distribution for the models with $M=10^{10}\Msun$ (upper), $M=10^{9}\Msun$ (middle), $M=10^{8}\Msun$ (lower). The \textbf{left panels} show the density distributions on the major axis the density information can be obtained from the axis ratios of Figure~\ref{triaxial}. The \textbf{right panels} show the accumulated mass inside the radius $r$. The dashed black lines in the \textbf{right panels} are defined as ${a_0 r^2\over G}$, which are the watersheds of enclosed mass producing MONDian dominating gravity (below the lines) and Newtonian dominating gravity (upon the lines). The colouring of the lines is representative of the evolutionary stage of the model and corresponds to Fig.~\ref{dmde}.}\label{mass}
\end{center}
\end{figure}

Due to parts of the kinetic energies spilling out of the cores (cf.
sub-section~\ref{energydistributions} and~\ref{kinetics}), the mass
densities could redistribute at the same time. Indeed, we find that
there are outflows of mass: the cuspy centers with an initial value of 
$\gamma=1$ are flattened. This can be viewed in Fig.~\ref{mass}
where we show the densities along the major axis (\textbf{left panels}) and cumulative (\textbf{right
panels}) mass distributions for our MOND models. With regards to the density panels, the three models show a similar behavior.
It is clear that the mass is redistributed during the relaxation with losses
in the very central region of $r<0.5$~kpc and gains outside. The density curves are ocsillating around the initial analytical density as given by Eq.~\ref{den}. Therefore,
the system becomes slightly less cuspy and keeps the triaxial density after reaching equilibrium. Note
that the density distribution still remains triaxial after the system
is in equilibrium, and there is no obvious evolution
within 200 Keplerian Dynamical times (at the typical scale a=1.0~kpc).
The right panels of Figure~\ref{mass} show the total mass inside the radial direction $r$. The black dashed straight lines in right panels are defined by $M_0={a_0r^2\over G}$, the mass to produce the gravity acceleration $a_0$ in a point mass approximation. $M_0$ is the watershed of the enclosed mass producing MOND and Newton dominating gravities. At a certain radius $r$, when the enclosed mass is smaller than $M_0$, there occurs a transition to MONDian gravity. We find that in all of the three models MONDian effects cannot be ignored. Even for $1.0\times 10^{10}\Msun$, the MONDian gravity dominates the regions of $r> 10^{0.3} \sim 2$~kpc. Obviously, the model $M=1.0\times 10^8 \Msun$ is in deep MOND region. The colours show the evolution of the systems.
Mass in the inner part of the system is lost during the density re-distribution,
while in the outer part, beyond 4~kpc, the total mass is not affected. 
We further note that after the redistribution (during the
relaxation) the mass distribution has stabilized.

\subsection{Shape}
\begin{figure}{}
\begin{center}
\resizebox{7.5cm}{!}{\includegraphics{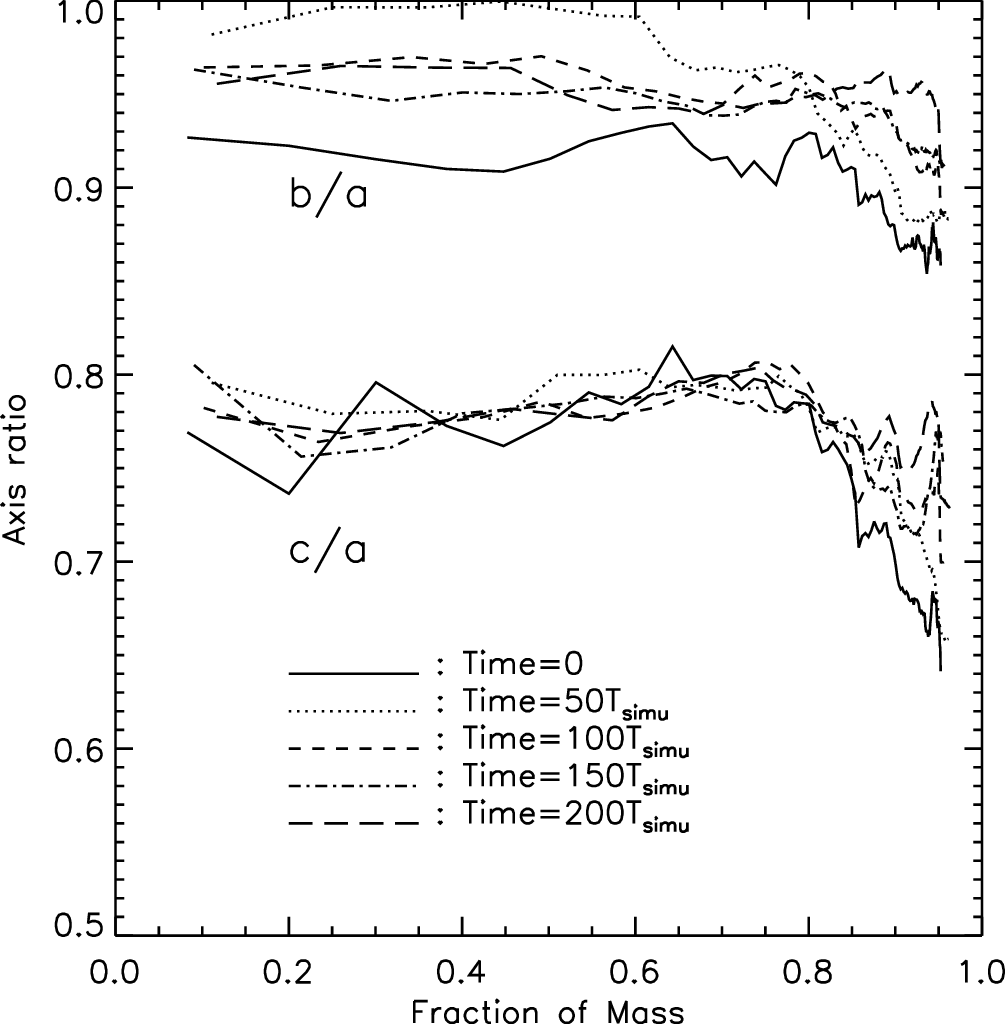}}\vskip 0.5cm
\resizebox{8.5cm}{!}{\includegraphics{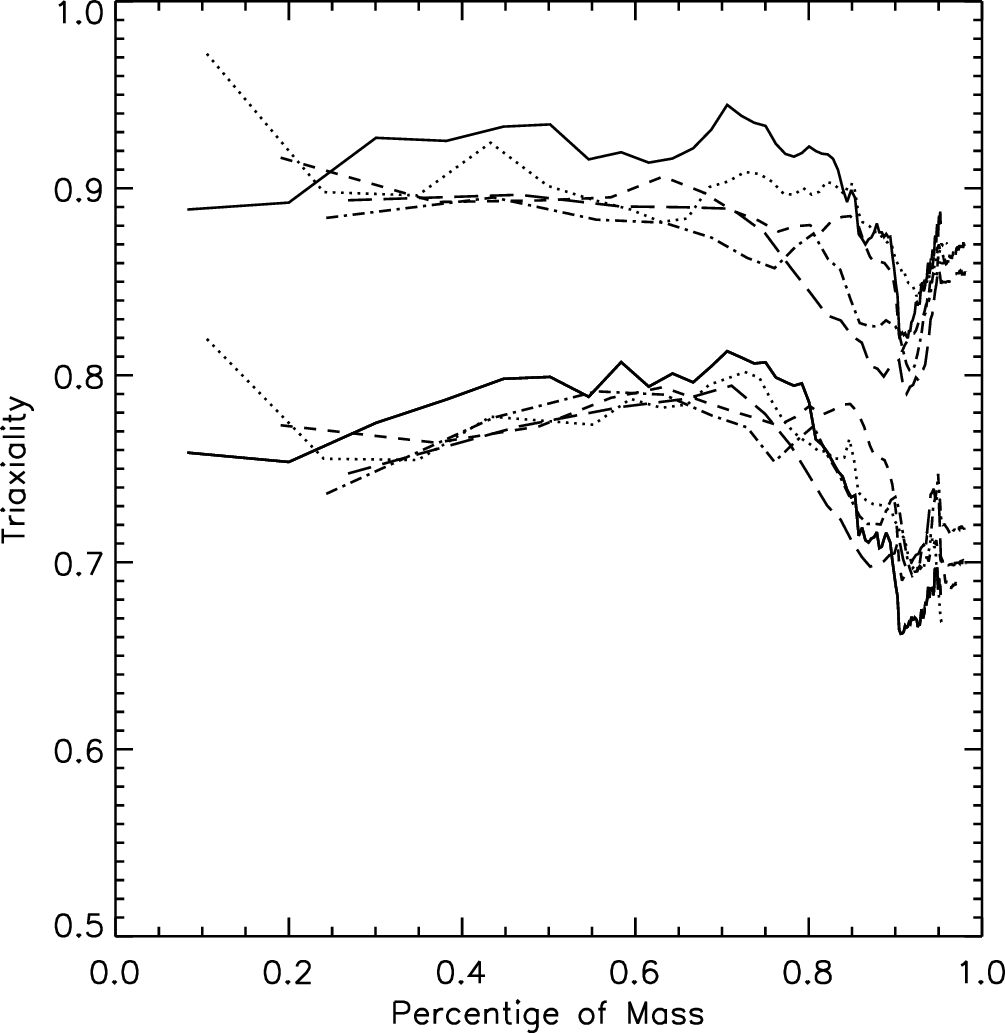}\includegraphics{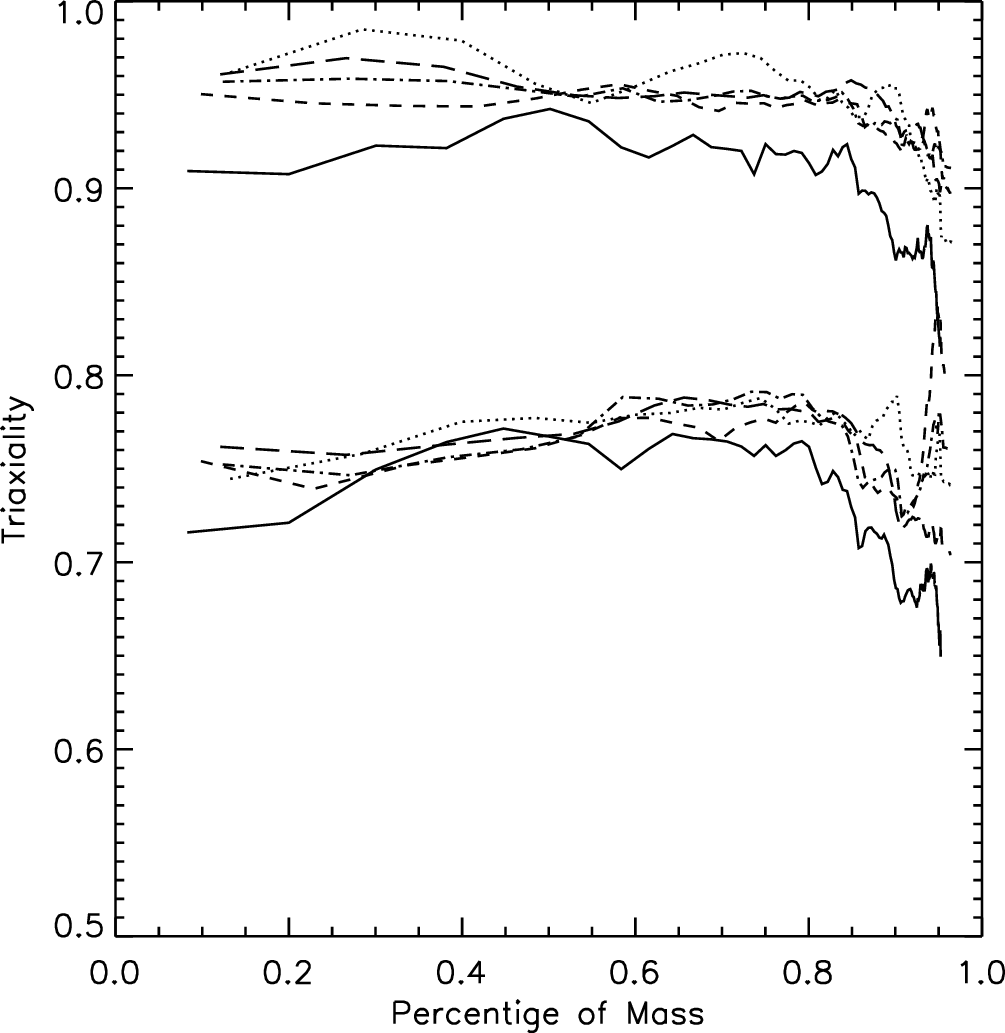}}
\vskip 0.5cm
\makeatletter\def\@captype{figure}\makeatother
\caption{Evolution of axis ratios with the median model of a total mass of $1.0\times 10^{9}\Msun$ (Upper panel), $1.0\times 10^{10}\Msun$ (lower left panel) and $1.0\times 10^8 \Msun$ (lower right panel). The lower and upper series of lines are for the ratios of minor : major axis and intermediate : major axis, i.e.,  $\sqrt{I_{zz}/I_{xx}}$ and $\sqrt{I_{yy}/ I_{xx}}$.  The different line symbols are defined the same as in the figure: solid, dotted, short dashed, dot-dashed and long dashed lines are for system evolving 0, 50, 100, 150 and 200 $T_{simu}$, where $T_{simu}$ is the Newtonian orbital time at 1.0 kpc .}\label{triaxial}
\end{center}

\end{figure}

\begin{figure}{}
\begin{center}
\resizebox{8.5cm}{!}{\includegraphics{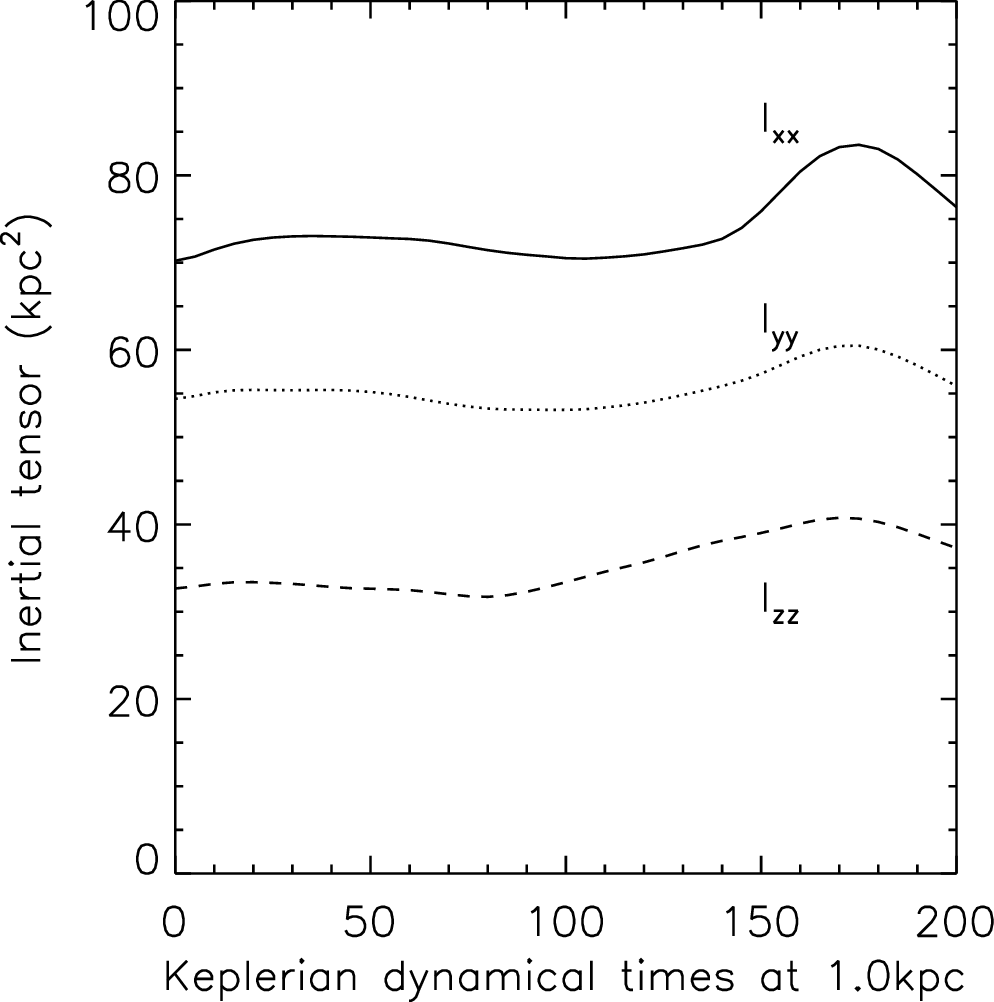}\includegraphics{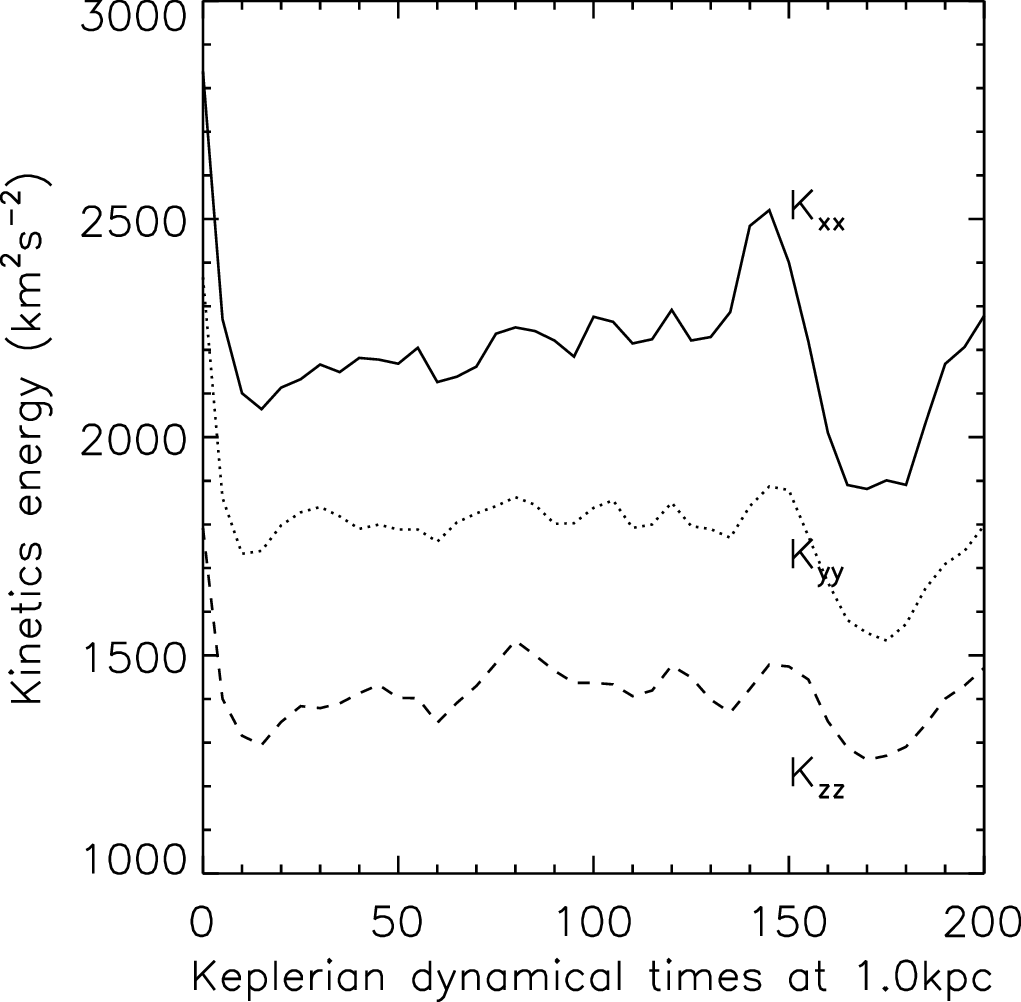}}
\resizebox{8.5cm}{!}{\includegraphics{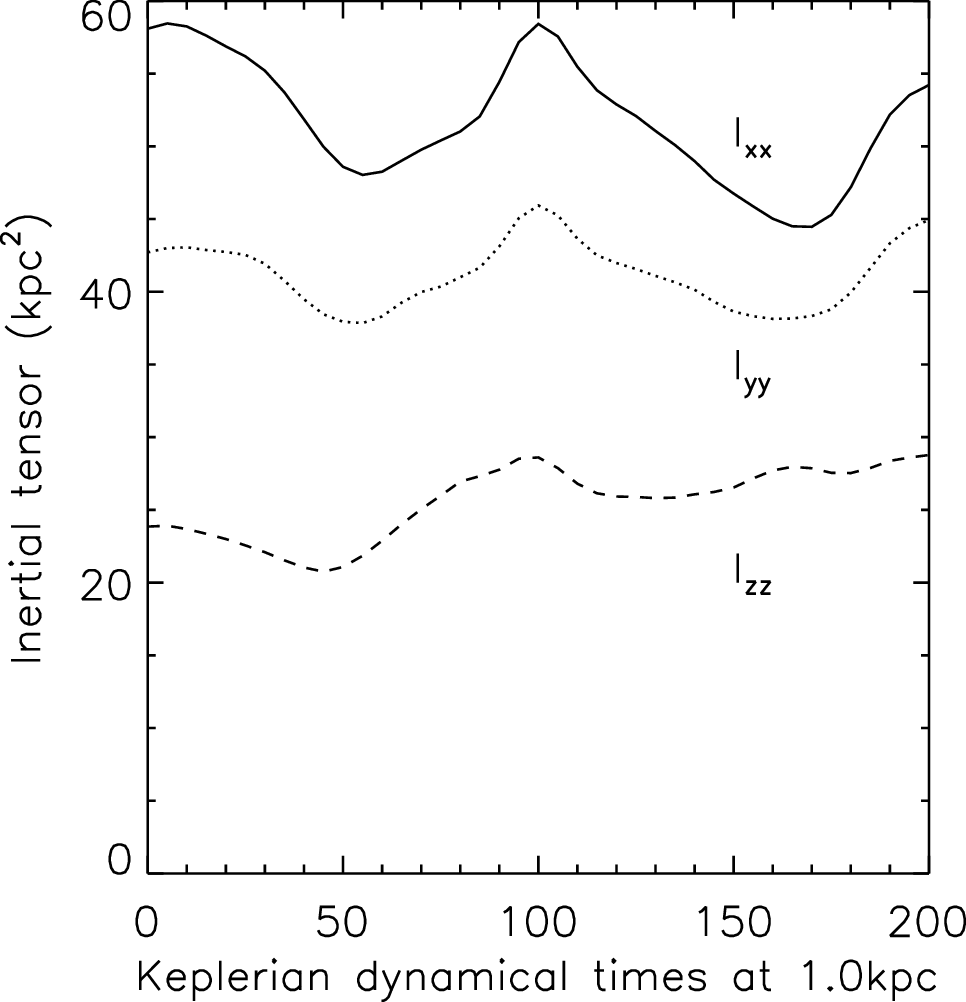}\includegraphics{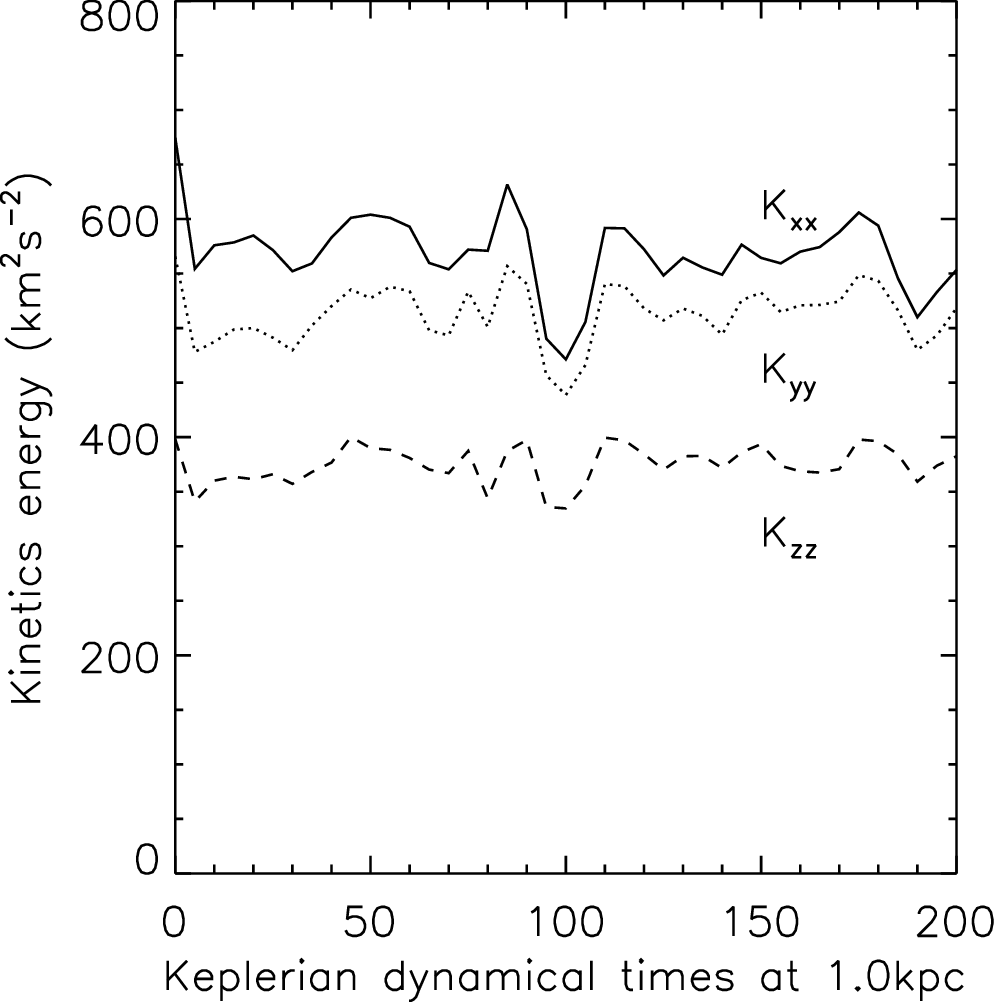}}
\resizebox{8.5cm}{!}{\includegraphics{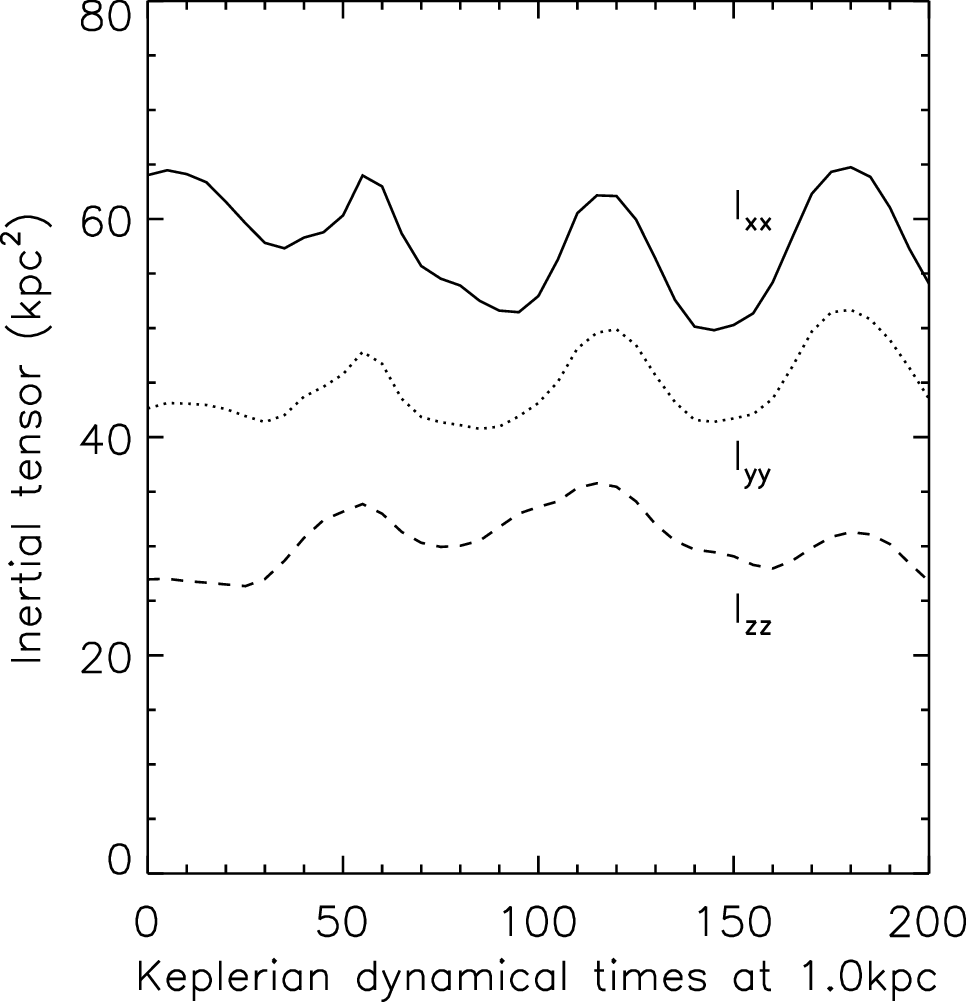}\includegraphics{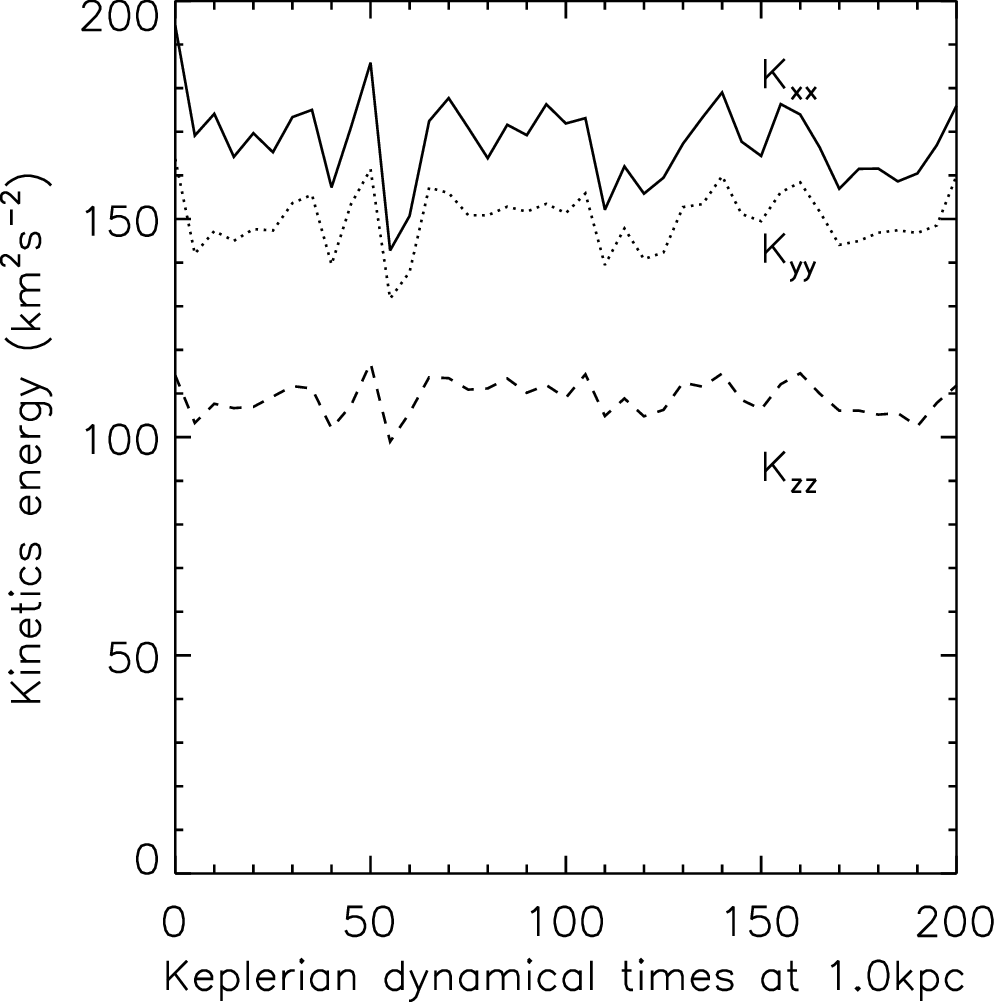}}
\vskip 0.5cm
\makeatletter\def\@captype{figure}\makeatother
\caption{\textbf{Upper}, \textbf{middle} and \textbf{lower panels} are different mass models of $10^{10}\Msun$, $10^9\Msun$ and $10^8\Msun$. The \textbf{left panels} are the evolution of the inertial tensor $I_{xx}$ (solid line), $I_{yy}$ (dotted) and $I_{zz}$ (dashed). The \textbf{right panels} are the evolution of kinetics energy $K_{xx}$ (solid), $K_{yy}$ (dotted) and $K_{zz}$ (dashed). The total simulation time is 200 $T_{simu}$.}\label{IxxKxx}
\end{center}
\end{figure}

As confirmed in \S \ref{massdistributions}, the mass redistributes inside our systems during relaxation. Hence, the question arises whether the shape (i.e., the initial triaxiality) remains stable or undergoes changes.
To address this, we show the evolution of the
axis ratios of the eigenvalues $\sqrt{I_{yy}/I_{xx}}$ and
$\sqrt{I_{zz}/ I_{xx}}$ of the moment of inertia tensor
$m_{ij}x_ix_j$, ($m_{ij}=M/N$) in Fig.~\ref{triaxial}. The three
models give similar results, hence we highlight the model with a
total mass of $10^{9}\msun$ in the upper panel. We find that the axial ratios (as a
function of radius) merely evolve about 10\% within 200 Keplerian
dynamical times (1.0 kpc). The system is rounder in the center, but
the whole system keeps the triaxial shape during the long stage of
evolution. It is clear that the axis ratio between the minor and major axes is more stable than that of the intermediate and major axes. Not only the ratios are almost constant, but also the
absolute values of $I_{xx}$, $I_{yy}$ and $I_{zz}$ seem in dynamic equilibrium and  stable, changing
 less than $20\%$ during the oscillation (cf. Fig~\ref{IxxKxx}). However, we
note that at the time $t=0$ the ratios of $\sqrt{I_{yy}/I_{xx}}$ and
$\sqrt{I_{zz}/ I_{xx}}$ do not accurately equal
$b:a=0.86$ and $c:a=0.7$ in most of the inner regions, which is caused by the numerical effects
in generating the $N$-body ICs. However, at the edge of the galaxy (i.e., including more than 80\% of the total mass), the axis ratios are close to the suggested ones. We conclude that the ICs generated by our
application of the Schwarzschild technique roughly lead to a 5\% error for
the axis ratios.

A study of the tensor kinetic energies $K_{xx}$, $K_{yy}$ and $K_{zz}$, defined as $K_{xx}=0.5<v_x\cdot v_x>$, shows a similar behaviour to the moment of inertia tensor analysis presented above. The ratios remain constant although the absolute values change by at most 20\%. This can again be verified in Figure~\ref{IxxKxx} where we plot the inertial (left panel) and kinetic energy (right panel) components for the three models. 


As a final note, considering existing Schwarzschild plus N-body simulations
in the literature, we find that the evolution seen in our MONDian cuspy
elliptical models is comparable to that seen in Fig.5 of Poon \&
Merritt (2004, ApJ 606, 774) for triaxial ellipticals in Newtonian gravity.
Our simulations are much longer than 10 crossing times, which provides a typical scale for
checking stability in Newtonian Schwarzschild simulations in the literature
(Poon \& Merritt 2004, Zhao 1996).

\section{Conclusions and Discussion}
We explored the stability and evolution of the triaxial Dehnen model (Dehnen 1993; Merritt \& Fridman 1996; Capuzzo-Dolcetta et al. 2007) with a $\gamma=1$ central cusp using MOND. We utilized the Schwarzschild method (Schwarzschild 1979) to build orbit models which were in turn used to generate initial conditions (ICs) for N-body simulations using the method outlined in Zhao (1996). These ICs were evolved forward in time for 200 Keplerian dynamical times (at the typical length scale of 1.0 kpc) by the numerical integrator NMODY developed by the Bologna group (Ciotti et al. 2006; Nipoti et al. 2007) and designed to include the effects of MOND. We additionally ran the same simulations with a second MONDian gravity solver \texttt{AMIGA} (Llinares, Knebe~\& Zhao 2008, cf. Appendix~\ref{sec:claudio}) based upon an entirely different grid-geometry to confirm the credibility of our results.

In our simulations, the virial theorem was satisfied at all times. We showed that the systems start in quasi-equilibrium with a short relaxation phase of approximately less than five Keplerian dynamical times (1.0 kpc). We found outflows of energy and mass from the centres of the systems under investigation. Hence, during the relaxation stage, there is a flattening of the initially present $\gamma=1$ cusp to a core. Despite the obvious mass redistribution, we need to acknowledge that the shape of the systems remained unchanged in the course of the simulations; the axis ratios of the eigenvalues of the moment of inertia tensor (as well as the kinetic energy tensor) stayed constant.

The effects of resolution of the simulations should not remain unmentioned. We found that the potential calculated from the N-body ICs differs by 10\% compared to the analytical potential. Furthermore, the analytically predicted velocity dispersions at the initial time are $107.3$~km/s, $54.2$~km/s and $29.3$~km/s for the models $M=10^{10}, 10^9, 10^8\Msun$, respectively. However, they do not match the (numerical) values plotted in right panel of Fig.~\ref{virial}. Hence we use the Clausius integral $|W|$ in Equation ~\ref{Clausius}, calculate the analytical densities for the systems at $t=0$, to minimize the errors. Moreover, we have found that due to the resolution limitation of the NMODY code, the errors accumulate during the simulations, which is insensitive to more massive systems, but becomes an issue when the mass of the system decreases. This causes small non-zero net velocities in the systems. The simulation centres of the systems with $10^9\Msun$ and $10^8\Msun$ move significantly after 200 Keplerian dynamical times (1.0 kpc) and hence we restrict our analysis to this time frame.
To further check the credibility of our results and the dependence on the code, we ran the simulations again with a technically substantially different code (\texttt{AMIGA}), which is also capable of integrating the analog to Poisson's equation (cf. equation~\ref{mond} in Appendix~\ref{sec:claudio}). The results are practically indistinguishable reassuring their tenability.


We like to close with the notation that our systems are isolated systems, corresponding to the cases of field galaxies. The self-potentials of the systems in MOND are logarithmic at large radii, therefore no stars can escape from such systems. However, for any system embedded in external fields, the potential is truncated when the strength of the external field becomes comparable to the internal field (Milgrom 1984; Wu et al. 2007). Therefore, Poisson's equation should be modified to
\beq
\nabla \cdot \left[\mu\left({|\nabla \Phi_{int}-\vec{g}_{ext}|\over a_0}\right)(\nabla \Phi_{int}-\vec{g}_{ext})\right] = 4\pi G\rho_b,
\eeq
where the $\mu$-function is determined by both the internal \textit{and} external gravitational accelerations. Hence the strong equivalence principle is violated, and the directions along and against the external field, the $\mu$-function has different values even though the mass density distributions are the same. A direct result is that the potentials become non-symmetric along and against the directions of the external field, i.e., a symmetric system is not in equilibrium due to the non-symmetry of self-potential. Therefore, MOND predicts that there are no real symmetric systems within the external gravity backgrounds. This will be explored in greater detail in a future paper (Wang et al. in preparation).

\section{acknowledgments}
We thank the anonymous referee for helpful suggestions and comments to the earlier version of the manuscript. We thank Luca Ciotti, Pasquale Londrillo, Carlo Nipoti for generously sharing their code, Martin Feix for polishing the English writing and Victor Debattista, Mordehai Milgrom and Christos Siopis for nice comments in the earlier version of the paper. We thank the Mordehai Milgrom and Francoise Combes for the comments to the paper. XW and HSZ acknowledges the Dark Cosmology Center of Copenhagen University and Sterrewacht of Leiden University. XW acknowledges the support of SUPA studentship. HSZ acknowledges partial support from UK PPARC Advanced Fellowship and National Natural Science Foundation of China (NSFC under grant No. 10428308). YGW acknowledges the support of the 973 Program (No.2007CB815402), the CAS Knowledge Innovation Program (Grant No. KJCX3-SYW-N2), and the NSFC grant 10503010. CLL and AK acknowledge funding by the DFG under grant KN 755/2. AK further acknowledges funding through the Emmy Noether programme of the DFG (KN 755/1).

\appendix
\section{Symmetry and Numerical Challenges} \label{numerics}
Evolving the systems without filtering the high frequent components of mass during the Legendre transformation in time for up to 200 Keplerian dynamical times (1.0 kpc) we find that they appear 'unstable'. However, a detailed investigation revealed that is a numerical effect rather than a physical instability. There is a small, uneven force in the $z$ direction which comes from the asymmetry of the density distribution of ICs for N-body and systems during the simulations 
and the errors obviously accumulate when the numbers of particles are not big enough. Further, the total momentum of the systems is not conserved giving a non-zero net velocity along the minor axis. Hence the code requires a large number of particles for smooth density distributions to make sure the tiny asymmetry of some particles does not affect the whole system.

\begin{figure}{}
\begin{center}
\resizebox{7.5cm}{!}{\includegraphics{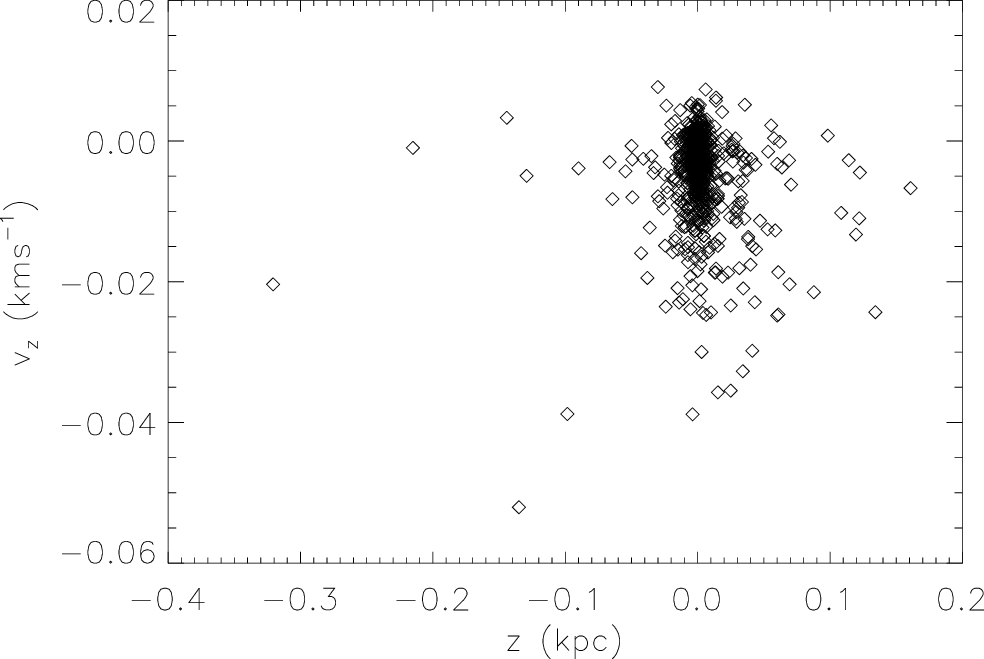}}
\vskip 0.5cm
\makeatletter\def\@captype{figure}\makeatother
\caption{The average values of positions ($\bar z$) vs. velocities ($\bar v_z$) projected on the z axis for each non-zero weight orbit. The model has the total mass of $10^9\Msun$.}\label{xvbar}
\end{center}
\end{figure}

Note that this effect is more serious when the systems are not symmetrized by utilizing 'mirror particles' inside the systems. It is known that particles generated from asymmetric orbits (e.g. the 'banana orbits') could break the symmetry of the systems in phase space. Furthermore, there are a couple of hundred of chaotic orbits with positive weights, and, during the Schwarzschild process, the time integration of 100 orbital times may not be long enough to obtain symmetry. We therefore show in Fig.~\ref{xvbar} the average value of positions and velocities along the $z$-axis for every orbit of the $N$-body ICs of the model with mass $10^9\Msun$. We plot $\bar z$ vs. $\bar v_z$ since the $z$-direction displays the most serious shifting. For a perfectly symmetric system the values of $\bar z$ and $\bar v_z$ should be close to zero, while we find that they are not.
Hence we need to symmetrize the systems prior to the N-body procedure. The simplest way to achieve this is by placing 'mirror particles' into the system, i.e. using a minus sign in front of the 6-dimensional components. Therefore, the total numbers of particles increases to $N\times 2^6=64N$.

\begin{figure}{}
\begin{center}
\resizebox{7.5cm}{!}{\includegraphics{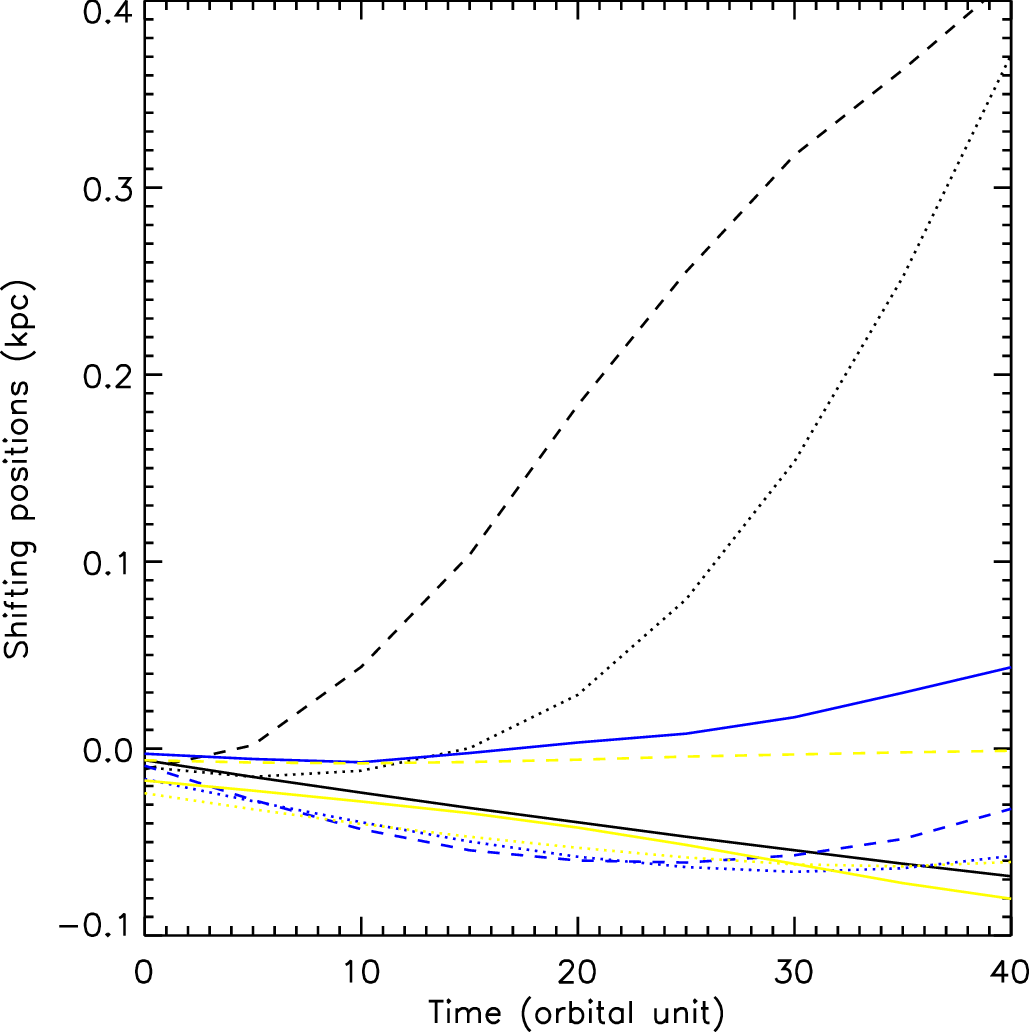},\includegraphics{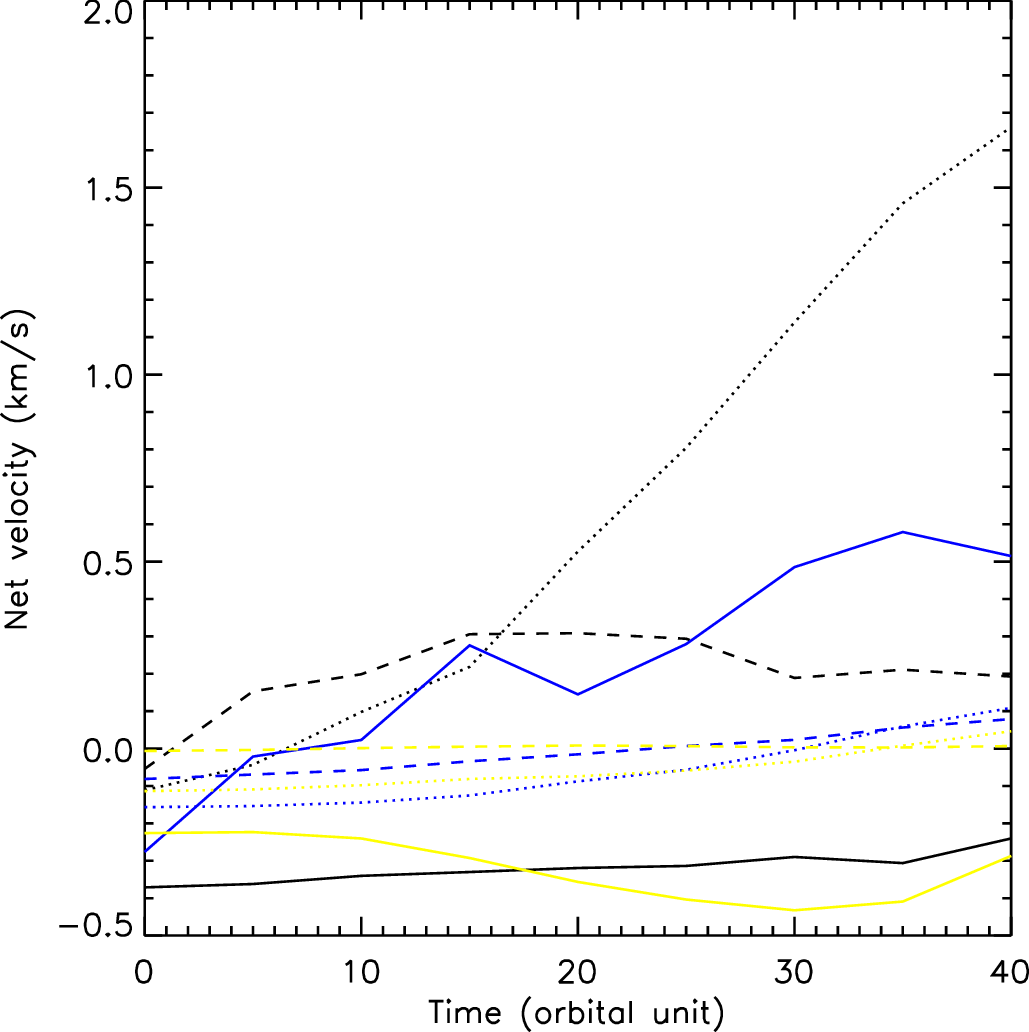}}
\vskip 0.5cm
\makeatletter\def\@captype{figure}\makeatother
\caption{Left panel: The shifted positions of the centre of mass of the three systems. The solid, dotted, dashed lines are for the systems with their total mass of $10^{10}\Msun$, $10^9\Msun$ and $10^8\Msun$. The colours of black, yellow and blue correspond to the z, y and x axis. The right panel: net velocities of the systems in the three axis directions. }\label{shifts}
\end{center}
\end{figure}

To illustrate (and quantify) this effect we plot in Fig.~\ref{shifts} the centre-shifts along the three axes (left panel) as well as the evolution of the net velocities along the same axes (right panel) during the first 40 Keplerian dynamical times (1.0 kpc) for the models without the 64 mirrors. We find that the  most massive system (i.e. $10^{10}\Msun$, which is in mild-MOND gravity) is least affected by these numerical artifacts. As a matter of fact, this particular system shows credible signs of stability even after 200 Keplerian dynamical times (1.0 kpc). We need to acknowledge that this is partly due to our definition of the time unit (cf. Eq.~\ref{time}): it is shorter for more massive systems. However, all the analysis presented in this paper indicates that the system is stable despite the apparent numerical artifacts of the code. We though cannot evolve the system further in time as the accumulation of errors would lead to substantial deviations from the system's equilibrium state; but this 'instability' is caused by numerics rather than physics!
Given the technical particulars of the NMODY code such a centre-shift will lead to a decrease in resolution as the code utilizes a spherical grid.

\section{Comparison with another MOND solver} \label{sec:claudio}
As just highlighted in Section~\ref{numerics}, there are numerical challenges to evolving our systems under MONDian gravity using the N-body code NMODY. In order to confirm that the results are not unique to this one code we therefore decided to also use another novel solver for the MONDian analog to Poisson's equation, namely the \texttt{AMIGA} code (Llinares, Knebe \& Zhao 2008). \texttt{AMIGA} is the successor to \texttt{MLAPM} (Knebe, Green \& Binney 2001) that has recently been adapted to also solve Eq.~\ref{mond}.\footnote{We like to note in passing that \texttt{MLAPM} has already been successfully applied to study cosmological structure formation under MOND (Knebe~\& Gibson 2004) under certain assumptions.} The code utilizes adaptive meshes in Cartesian coordinates in a cubical volume as opposed to the spherical grid of NMODY. The solution is obtained via multi-grid relaxation and we refer the interested reader to Llinares et al. (2008) for more details.

However, here we need to elaborate upon the boundary constraints as we cannot assume that the potential on the boundary will be a constant: the box is a cube and not a sphere. We decided to use the solution for a point mass in the center of the box



\beq
\Phi(r) = -\frac{GM}{2}\left( \frac{1}{r} - \frac{1}{r_0}\right) +
\left( f(r) - f(r_0) \right),
\eeq
with
\begin{equation}
\begin{array}{rcl}
f(r) & = & \displaystyle -\sqrt{GMa_0}\\
& & \displaystyle \left[\frac{-1}{2r} \sqrt{q^2 + 4r^2} +
 \ln\left(2r +  \sqrt{q^2+4r^2} \right) \right]\\
 \\
q^2 & = & \displaystyle \frac{GM}{a_0},
\end{array}
\end{equation}

\noindent
where, $M$ is the total mass in the box and $r_0$ is a length scale (a constant of integration).  For $a_0\rightarrow 0$ we recover the Newtonian solution and for $a_0$ finite and $r\rightarrow\infty$ we have $\ln(r)$, which is the typical behaviour for any MONDian solution.  In the case that we use $r_0=B$, with $B$ being the size of the cubical box, we end up with $\Phi=0$ in a sphere of radius $B$, that is equivalent to the conditions used in NMODY.

We now run simulations with the same Initial Conditions for N-body as used with NMODY utilizing a domain grid with $128^3$ cells. Each of these domain grid cells is refined and split into eight sub-cells once the number of particles inside that cell is in excess of 6.  The box size is $B=165.5152$kpc and the scale for the boundary conditions is $r_0=82.7576$ Kpc (half of the box).

The results obtained are similar to the NMODY simulations.  The system is stable with a normal secular evolution.  
We observe the same kind of evolution.
We confirm that all other quantities behave in a similar manner too, and hence are confident that the results presented in the previous section~\ref{simulations} are not dominated and/or contaminated by numerical artifacts.


\section{Longer time evolution of the $10^{10}\Msun$ model}
We initially ran the models for 200 simulation time units (i.e. 200 circular orbital times at the length of 1.0 kpc, see Table~\ref{TotalTimes}). Hence the most massive model has been evolved for the least time, about 1 Gyr. For brighter galaxies because they are also bigger galaxies, the dynamical time is longer. This is the opposite to our trend in Table 1.
This can be understood because our toy galaxies do not sit on the fundamental
plane.  To see this, we estimate $r_h$, the  characteristic length of an elliptical galaxy on the
fundamental plane
(e.g., Eq. 9 in Zhao, Xu, Dobbs 2008, which follows Faber et al. 1997 ).
\beq
\log {GMr_h^{-2}\over 350 \times 10^{-10}m/s^2}=-1.52\log {M\over 1.5\times 10^{11}\Msun}\pm0.5,
\eeq
we find $r_h=0.082^{+0.049}_{-0.032}$~kpc for the model with $10^{10}\Msun$. This is one order of magnitude smaller than our assumed size $1$~kpc.
The dynamical time for a $10^{10}\Msun$ galaxy sitting on the fundamental plane is about 0.1 Million years.

\begin{figure}{}
\begin{center}
\resizebox{7.5cm}{!}{\includegraphics{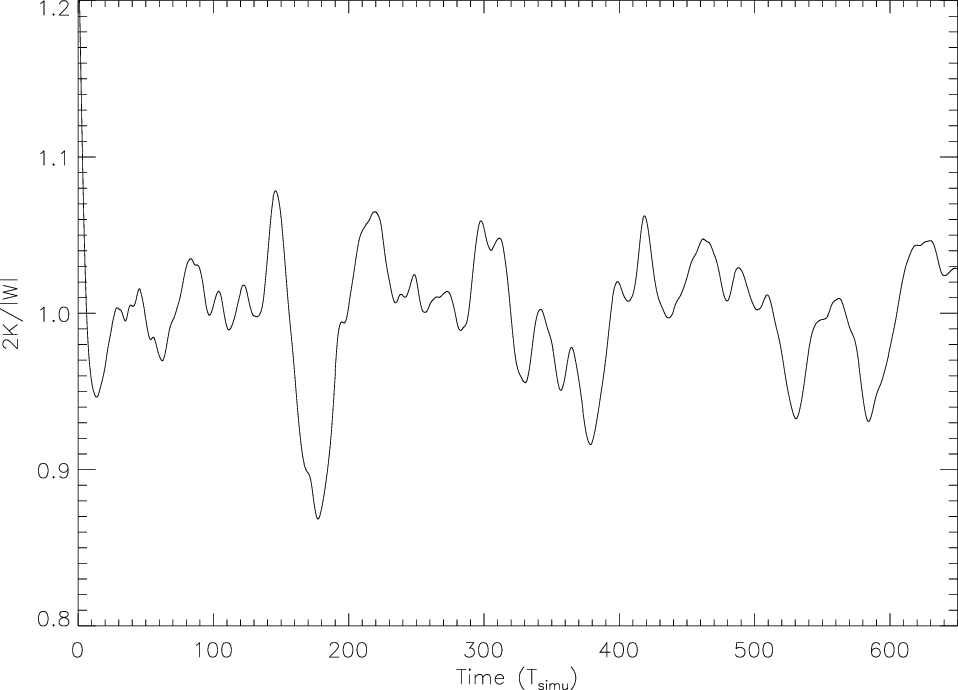}}
\vskip 0.5cm
\makeatletter\def\@captype{figure}\makeatother
\caption{The virial ratio of model with $10^{10}\Msun$, simulating 3 Gyrs (650 circular orbital time at typical length of 1.0kpc). }\label{longvirial}
\end{center}
\end{figure}

To be on the safe side, we re-ran the model for about 3~Gyrs ($650T_{simu}$), and show its virial ratio (Fig.~\ref{longvirial}). And our conclusion in the \S \ref{virialratio} does not change.  The virial ratio
oscillates around 1 within 10\% at most.

\end{document}